\documentclass[reprint,showpacs,preprintnumbers,amsmath,amssymb,showkeys,pre]{revtex4}

\usepackage{graphicx}
\usepackage{dcolumn}
\usepackage{bm}
\usepackage{amsmath}
\usepackage{amssymb}
\usepackage{amsfonts}
\usepackage{color}
\usepackage{dsfont}
\usepackage{epsfig}
\usepackage{hyperref}
\usepackage{mathrsfs}
\usepackage{multirow}
\usepackage{pifont}
\usepackage[T1]{fontenc}
\usepackage{relsize}
\usepackage{graphicx}
\usepackage{wasysym}
\usepackage{txfonts}   
\usepackage{subeq}
\usepackage{slashbox}
\usepackage[normalem]{ulem}
\usepackage[usenames,dvipsnames]{xcolor}

\newcommand{\sgn}{\text{sgn}}
\def \gv#1{\mbox{\boldmath $#1$}}
\renewcommand \vec \gv

\graphicspath{ {./Figures/},{./} }

\begin{document}

\title{Mechanical energy dissipation induced by sloshing and wave breaking in a fully coupled angular motion system. Part II: Experimental Investigation.} 

\author{B. Bouscasse}
\email{benjamin.bouscasse@cnr.it}
\affiliation{CNR-INSEAN \\ \mbox{Marine Technology Research Institute, Rome, Italy}}
\affiliation{Aeronautics Department (ETSIA), \mbox{Technical University of Madrid (UPM), 28040 Madrid, Spain}}
\author{A. Colagrosssi}
\email{andrea.colagrossi@cnr.it}
\affiliation{CNR-INSEAN \\ \mbox{Marine Technology Research Institute, Rome, Italy}}
\author{A. Souto-Iglesias}
\email{antonio.souto@upm.es}
\affiliation{Naval Architecture Department (ETSIN), \mbox{Technical University of Madrid (UPM), 28040 Madrid, Spain}}
\author{J. L. Cercos-Pita}
\email{jl.cercos@upm.es}
\affiliation{Naval Architecture Department (ETSIN), \mbox{Technical University of Madrid (UPM), 28040 Madrid, Spain}}

\date{\today}

\begin{abstract}
In Part I of this paper series, a theoretical and numerical model for a a driven pendulum filled with liquid was developed.
The system was analyzed in the framework of tuned liquid dampers (TLD) and
hybrid mass liquid dampers (HMLD) theory. In this Part II, in order to measure the energy dissipation resulting from shallow water sloshing,
an experimental investigation is conducted. Accurate evaluations of energy transfers are obtained through the recorded kinematics of the system.
A set of experiments is conducted with three different liquids: water, sunflower oil and glycerine. Coherently with the results of Part I, the energy dissipation obtained when the tank is filled with water can mainly be explained by the breaking waves. For all three liquids, the effects of varying the external excitation amplitude are discussed.
\end{abstract}
\pacs{47.11.-j, 47.15.-x, 47.10.ad}
\keywords{sloshing flows, dissipation, breaking waves, viscous effects, TLD, Tuned Liquid Dampers, Smoothed Particle Hydrodynamics, shallow water}
\maketitle
\section{Introduction}
A theoretical and numerical model for a fully coupled dynamical system, called hereinafter Pendulum-TLD, has been developed
in Part I \citep{bouscasse2013mechanical_partI_ARXIV} of the present work.
The mechanical system is essentially a non-linear driven pendulum, where the pendulum is a rectangular tank rotating around a fixed pivot.
The tank, partially filled with a liquid, receives torque from a mass sliding along
a linear guide fixed to the tank. This torque is the external excitation for the system motion.
Analogies with a tuned liquid damper (TLD) and
a hybrid mass liquid dampers (HMLD) have been provided in Part I.
in order to validate the models developed in
Part I and to provide insight into energy dissipation mechanisms in sloshing dampers with a focus on wave breaking influence, an experimental investigation is conducted in this Part II .

Much work has been done in order to assess the sloshing related loads that occur during forced harmonic motions (see e.g. \citet{bass1998,souto2006,diebold_etal_isope_2011_irregular_exc_forces}).
Concerning the coupling between a mechanical system and fluid dynamics, one of the most interesting experimental work is due to \citet{cooker1994water} who carried out decay experiments with a free oscillating tank suspended as a bifilar pendulum in the shallow-water limit. The most attractive feature of this system is that friction effects are negligible. \citet{herczynski2012experiments} also tested several shapes for a free swaying tank.
\citet{Marsh2011_jsv,So_semercigil_jsv2004_egg} present other decay tests with egg shaped forms, claiming that they are close to optimum in motion dampening for a range of filling heights.

Experiments for coupled roll motion in waves are found in e.g. \citet{armenio1996b,Nasar_etal_rollcoupled_oe2010}.
The effects of screens and baffles placed inside the tank, have been studied experimentally in the works of e.g. \citep{Tait_etal_2005_jfs_screens_tld,Tait2008,Firoozkoohi_Faltinsen_isope2010}.
The extra dissipation is achieved when viscous boundary layers separate from the many solid edges of a screen. There is a rapid generation of vorticity, and therefore an associated high rate of dissipation of mechanical energy by the viscous forces.
\citet{Pirner2007} carried out experimental tests to assess the efficiency of a rolling and translating tank to dampen external vibrations. They performed a frequency analysis of the device, testing different filling levels and liquids with different viscosities.

In the present paper, the Pendulum-TLD, described in Part I \citep{bouscasse2013mechanical_partI_ARXIV} and in \citep{bulian_etal_jhr09}, is experimentally studied. Only one frequency of excitation is considered, forcing the system to roll at the mechanical linear resonant frequency (see discussion in Part I). In the designed system, forces and energy are retrieved from kinematic quantities. This avoids the use of sensors for the measurement of force and torque which introduces practical difficulties and measurement uncertainties.
The accuracy of these results could make this work interesting for the validation of numerical models as well as to provide useful information for TLD or HMLD designers.

%

In order to investigate viscosity effects, three fluids are considered: water, sunflower oil and glycerine. They have a similar density but a very different viscosity.
Using these three liquids makes possible the investigation of the influence of Reynolds number for both the time-periodic state dissipation and the transient dynamics of the motion. Particular attention is given to the dissipation linked to wave breaking.

This Part II paper is organized as follows: the second section is dedicated to the experimental setup,
the third section presents experimental results for the empty tank, and in the fourth
the results for water are discussed. This is a relevant case since it is the
closest to an inviscid case, for which a theoretical model was developed in Part I.
In the fifth section, results for more viscous liquids are analyzed.
A summary of the energy dissipation measurements is presented considering the scales developed in Part I. Finally, a novel damping device, the hybrid pendulum mass liquid damper (HPMLD), is discussed before presenting conclusions.

This case is a benchmark for the SPHERIC SPH community 
and all data is available for reanalysis by third parties. 
See supplemental material at [URL will be inserted by AIP] for videos of the experiments.
Watching them across together with reading the paper is recommendable.
%
\section{Experimental setup}\label{sec:exp_setup}
\begin{figure}[t!]
\centering
\includegraphics[width=0.35\textwidth]{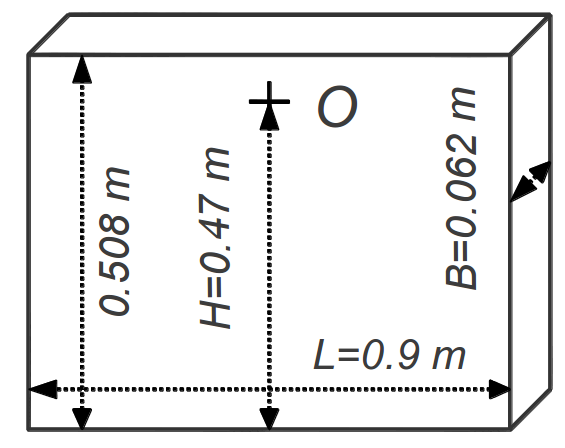} \hspace*{1.0cm}
\includegraphics[width=0.35\textwidth]{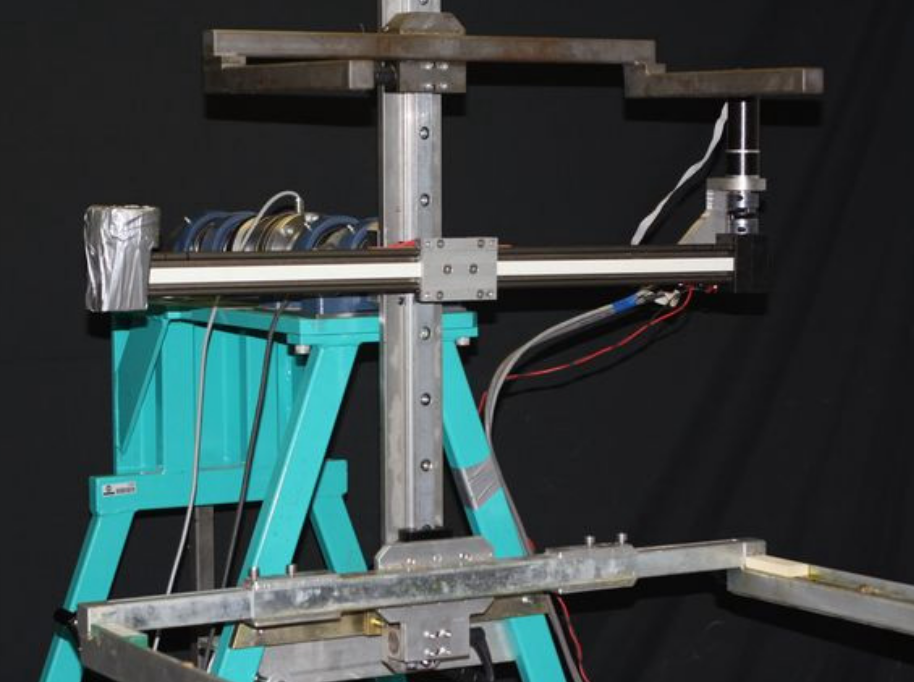}
\caption{ Tank dimensions and sliding device}
\label{fig:Tank1}
\end{figure}

The experiments were conducted with the tank testing device of the CEHINAV - UPM research group.
It is a single degree of freedom angular motion sloshing rig used for a number of experimental campaigns documented in the literature,
e.g. \citep{delorme_colagrossi_etal09,degroote_etal_cmame_2010,brizzolara_etal_2011_saos,idelsohn_etal_09}
and described thoroughly in \citep{souto_botia_martin_perezarribas_part0_oe2011}.
The tank chosen for the simulations is depicted with its dimensions
in Fig. \ref{fig:Tank1}. In particular, the tank length $L$ is 0.90 m, and the width $B$ is 0.062 m.
The length $l=0.1$ m is defined as a characteristic length of the system,
indeed, the filling height adopted will be of this magnitude (almost shallow water
regime) as well as the amplitude of the sliding mass motion.
The distance $H$ between the center of rotation and the tank bottom is set equal to 0.47 m.
The rotation center is above the center of gravity of the whole system,
implying that the system is stable at the equilibrium position.

The tank is intentionally narrow along the z-direction, i.e. the direction
perpendicular to the paper, in comparison with the horizontal and
vertical dimensions, in order to have a predominantly two
dimensional flow. A 0.60 m long linear guide is mounted parallel to the tank bottom passing though the rotation center. A controllable
electrical engine slides a mass, $m = 4.978\,\mathrm{kg}$, along the guide with a specified motion.
The weight and inertia of the mass generates a torque that acts as the external excitation to the system, and gives energy to the tank, which begins to roll.
Consequently, sloshing is induced and influences the dynamics of the rigid body angular motion.

The sloshing tank rotates on a fixed
plane around the fixed pivot $\vec O$. The moment of inertia around $\vec O$ is $I_0 = 26.9$
$\mathrm{kg}.\mathrm{m}^2$ and the static moment of the rigid system around $\vec O$ is $S_G = m_{tank} \eta_G = - 29.2 \mathrm{kg}.\mathrm{m}$.
The sliding mass moves with a defined harmonic motion $\xi_m (t)$,
\[
\xi_m(t)=A_m\,\sin(\omega\,t)
\]
where $A_m$ is the amplitude of the mass oscillation and $\omega$ is the oscillation frequency. $T$ is the related period set equal to the resonance period of the mechanical system $T_1 = 1.925\, \mathrm{s}$.
The sliding mass motion amplitude $A_m$ is set to $0.05$, $0.10$, $0.15$ and $0.20$ m.
Since $\xi_m$ is imposed, the state of the dynamical system can be defined as
a function of the angle $\phi$ and its derivatives.

\begin{figure}[ht]
\centering
\includegraphics[width=0.70\textwidth]{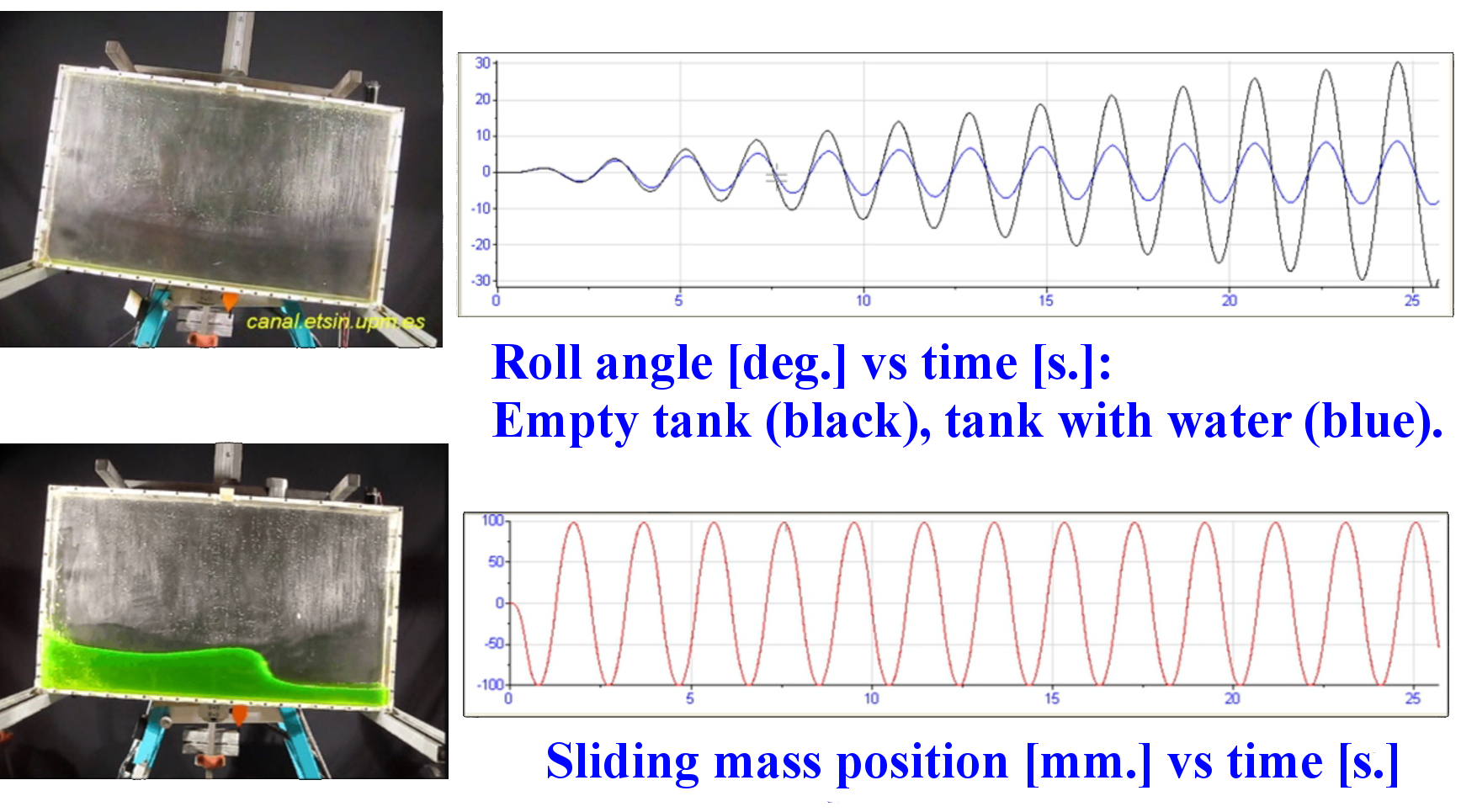}
\caption{Experimental device in action, top: empty tank, bottom: tank partially filled with water. Videos available as supplementary materials [URL will be inserted by AIP]
}
\label{fig:supp}
\end{figure}

In order to understand how this system works, the reader is referred to the video presented as supplementary material for Fig. \ref{fig:supp}. On the left part of this figure, sample cases with the empty tank and the tank partially filled with water are presented. In the top panel the angular motion of the tank is presented for these two cases. The influence of the fluid on the tank motion time history can be appreciated. Finally, in the bottom panel, the motion of the exciting sliding mass, which is the same for both cases, is plotted.

\section{Dynamics of the system with the empty tank}\label{empty_tank}

As shown in Part I, the dynamics of the system with the empty tank is not a straightforward problem due to the strong non-linear features. The analysis of Part I focused mainly on the time-periodic state, and results were obtained through a numerical integration of equation (III.22)
in Part I. The transient dynamics was shown to lead to very large angles.

\begin{figure}[ht]
\centering
\includegraphics[width=0.8\textwidth]{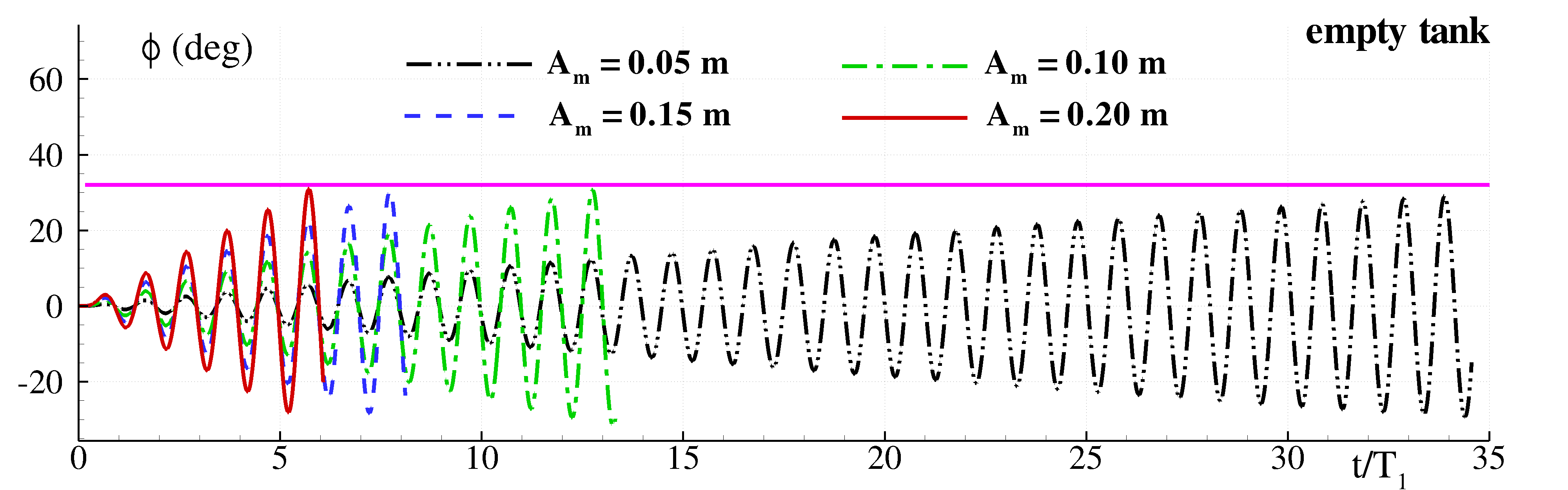}
\caption{Empty tank: experimental roll angle $\phi$ plotted as a function of time for different $A_m$ value using the excitation frequency $\omega\,=\,\omega_1$.
The horizontal purple line indicates the safety margins of the sloshing rig ($35$ degrees).}
\label{fig:Rollangle_empty}
\end{figure}

In order to remain within the safety operational margins of the sloshing rig, the experiments are stopped when the roll angle $\phi$ exceeds $35^\circ$
(see Fig. \ref{fig:Rollangle_empty}). This constraint impedes the time-periodic state showed in Part I to be reached. 
In the early phase of the test, $\Delta E_{mass/tank}$ is converted in $[E_{tank}^{mech}]_t^{t+T}$ and partially in $\Delta E_{friction}$. The system can be said to behave linearly with $A_m$ in that early phase.

\begin{figure}[ht!]
\centering
\includegraphics[width=0.8\textwidth]{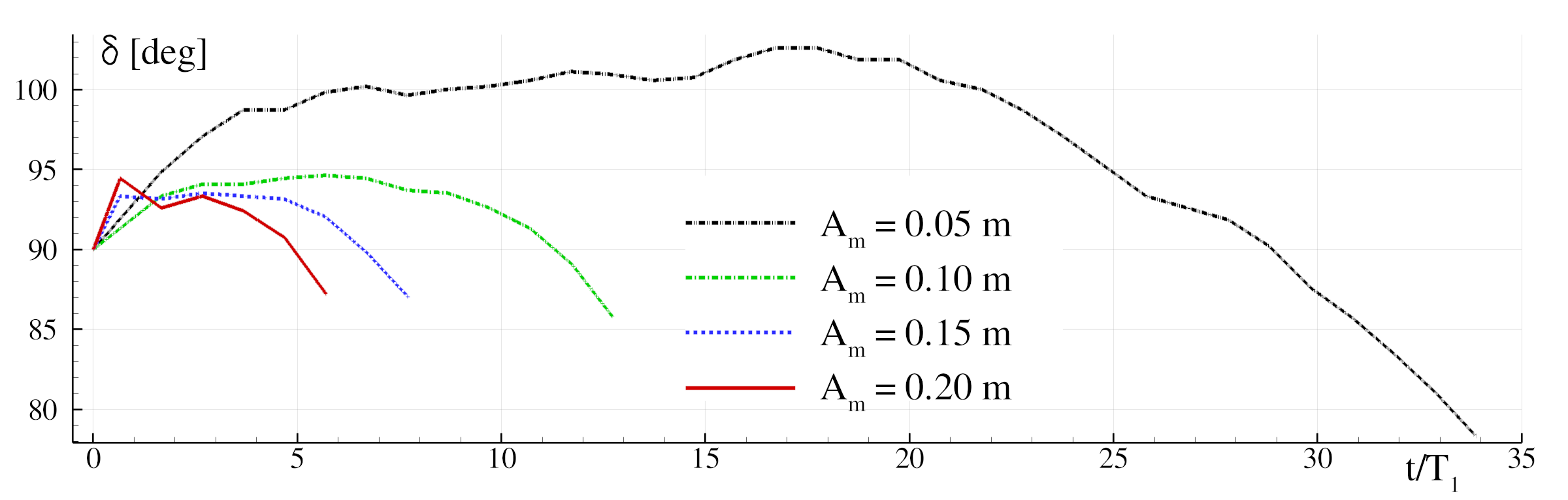}
\caption{Empty tank: shift function $\delta$ obtained from experiments plotted as a function of time for different $A_m$ value using the excitation frequency $\omega\,=\,\omega_1^m$.}
\label{fig:delta_empty}
\end{figure}

The primarily linear behavior seen on the experimental plots is confirmed in Fig. \ref{fig:delta_empty}, where
the lag function $\delta(t)$ (see definition in Part I section II.C) is plotted as a function of time.
$\delta$ remains close to $90^\circ$ (meaning that the roll motion is in quadrature with the sliding mass) for all amplitudes $A_m$.


\section{Analysis of the dynamical system: tank filled with water.}\label{ss:water}

In this section, the analysis of the dynamical system is considered when the tank is partially filled with water. The filling level is set equal to $h=0.092$ m and has been chosen so that the first sloshing period matches the resonance period of the structure $T_1=1.925$ s (see Part I). The period for the motion of the sliding mass $T$ has been set equal to $T_1$, and the excitation amplitudes $A_m$ are again $0.05\,,0.10\,,0.15\,,0.20\,m$, labeled respectively  as Series 1,2,3,4.

\begin{figure}[ht!]
\centering
\includegraphics[width=0.24\textwidth]{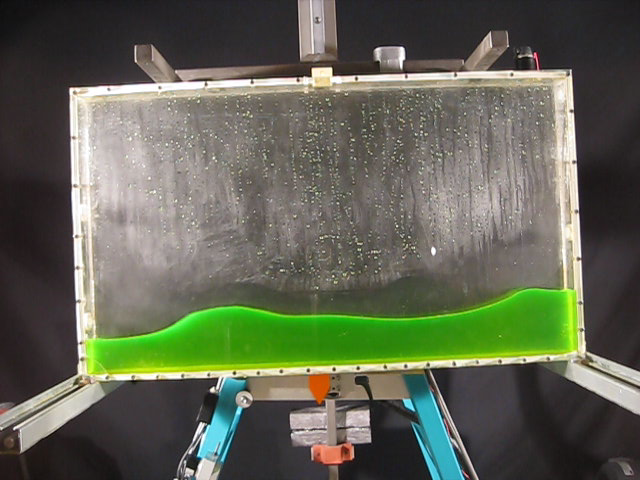}
\includegraphics[width=0.24\textwidth]{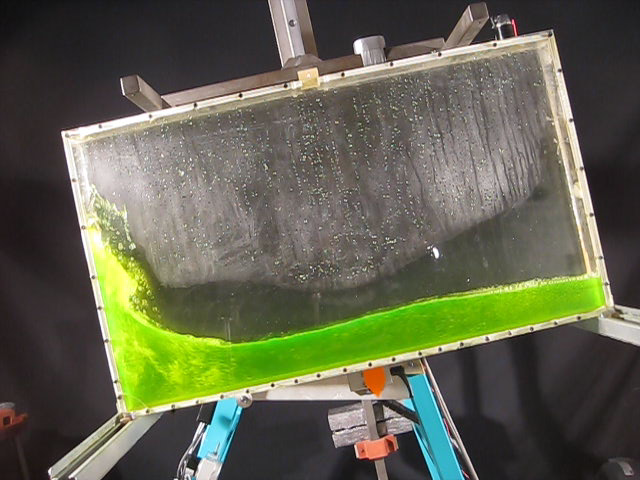}
\includegraphics[width=0.24\textwidth]{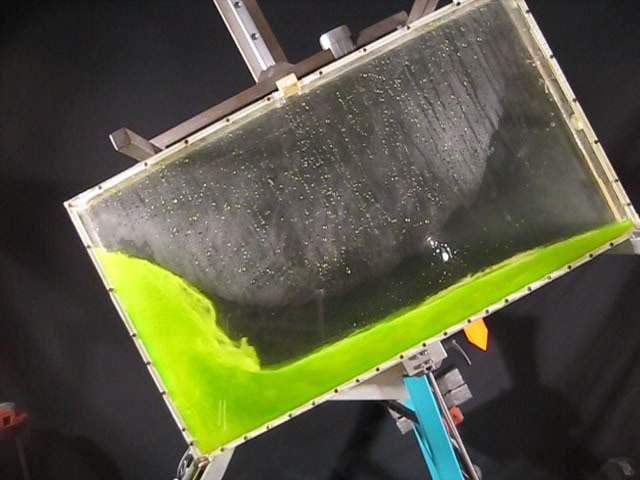}
\includegraphics[width=0.24\textwidth]{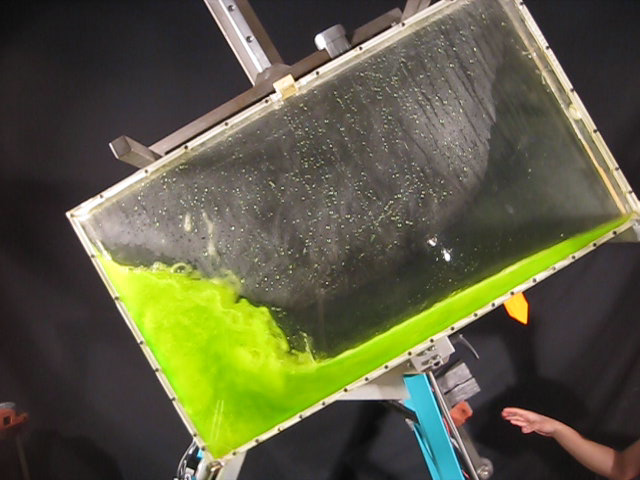}
\includegraphics[width=0.24\textwidth]{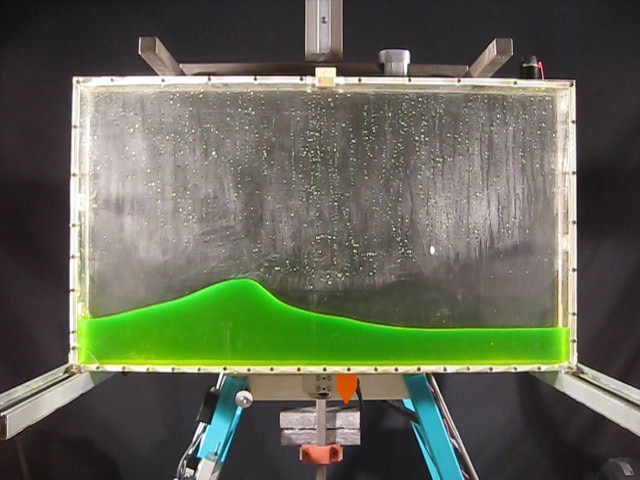}
\includegraphics[width=0.24\textwidth]{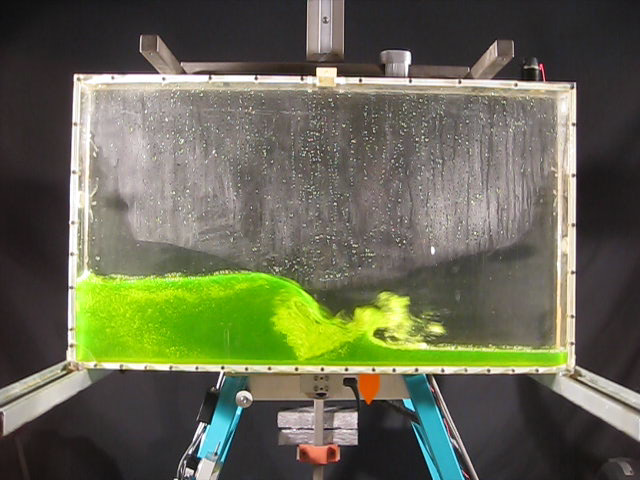}
\includegraphics[width=0.24\textwidth]{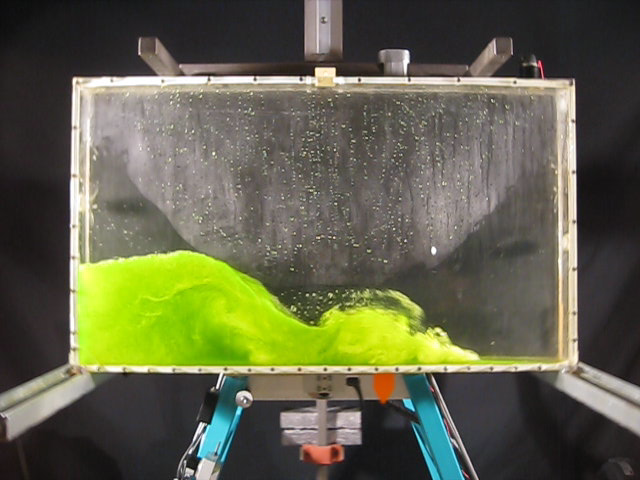}
\includegraphics[width=0.24\textwidth]{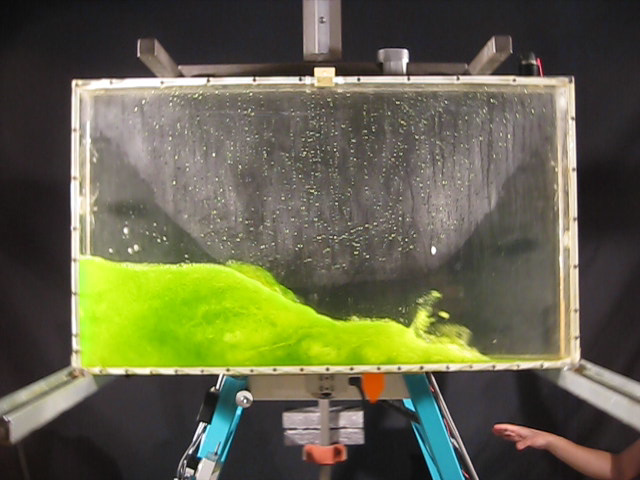}
\caption{Sloshing wave in tank filled with water from left to right $A_m =$  $0.05, 0.1, 0.15, 0.2$ m. Videos available from
[URL will be inserted by AIP]. 
} \label{fig:EvolSlosh_Water_allA}
\end{figure}

A discussion is presented for the four excitation amplitudes since distinct features in the sloshing regimes are observed; these are summarized in Table \ref{tab:sloshing_regime}.
In order to illustrate them, some pictures of the flow are presented in Fig.
\ref{fig:EvolSlosh_Water_allA} for each of the Series. Two instants are selected for each case, one with
the maximum angle and one with a flat angle.

\begin{table}[ht!]
\begin{center}
\vspace*{0.2cm}
\begin{tabular}{|c||c|c|c|c|}
\hline
                     &$A_m$ (m) & motion regime   & developed flow \\
                     \hline
\hline Series $1$  & $0.05$  & weak        & wave train without breaking event            \\[2pt]
\hline Series $2$  & $0.10$  & moderate    & plunging breaker in the middle of the tank  \\[2pt]
\hline Series $3$  & $0.15$  & strong      & strong hydraulic jump flow                   \\[2pt]
\hline Series $4$  & $0.20$  & very strong & an almost dam-break flow                     \\[2pt]
\hline
\end{tabular}
\vspace*{0.2cm}
{\footnotesize \caption{\label{tab:sloshing_regime} Sloshing regime induced in the oscillating
tank when using water.}}
\end{center}
\end{table}

For the first Series, only a wave train without breaking events takes place. For the second Series,
a plunge breaking event develops towards the middle of the tank.
For the most violent case, Series 4, when the maximum roll angle is reached, almost
all the water accumulates on the tank side and a quasi dam-break type flow occurs.
\subsection{Roll angle}
\begin{figure}[h!]
\centering
\includegraphics[width=0.84\textwidth]{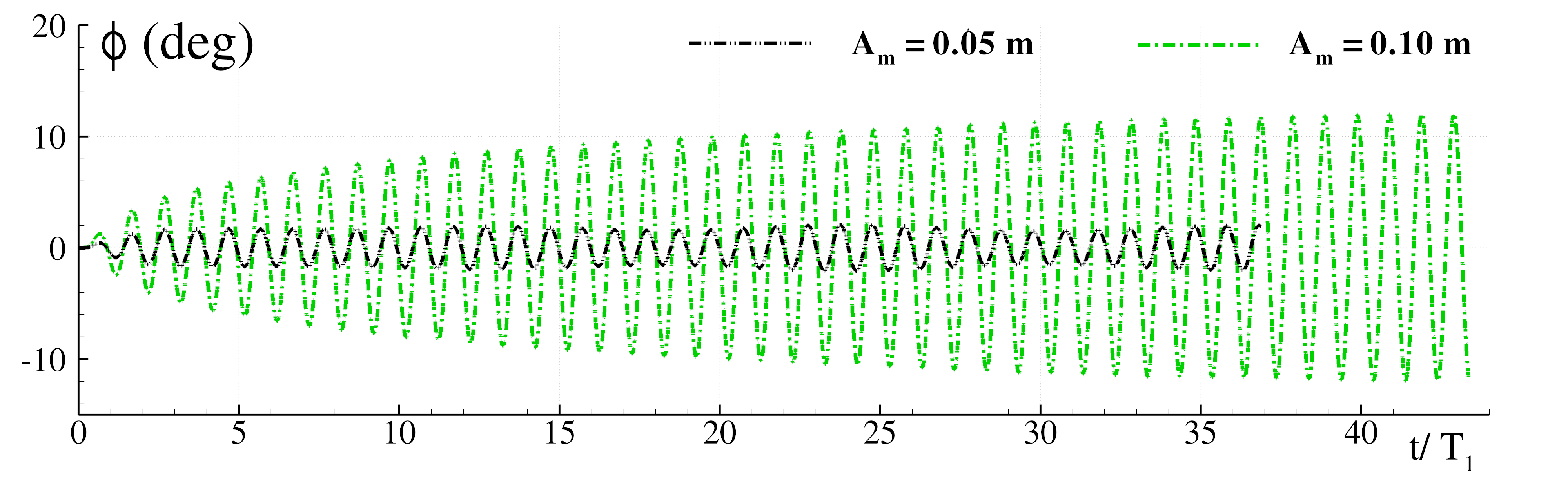}
\includegraphics[width=0.84\textwidth]{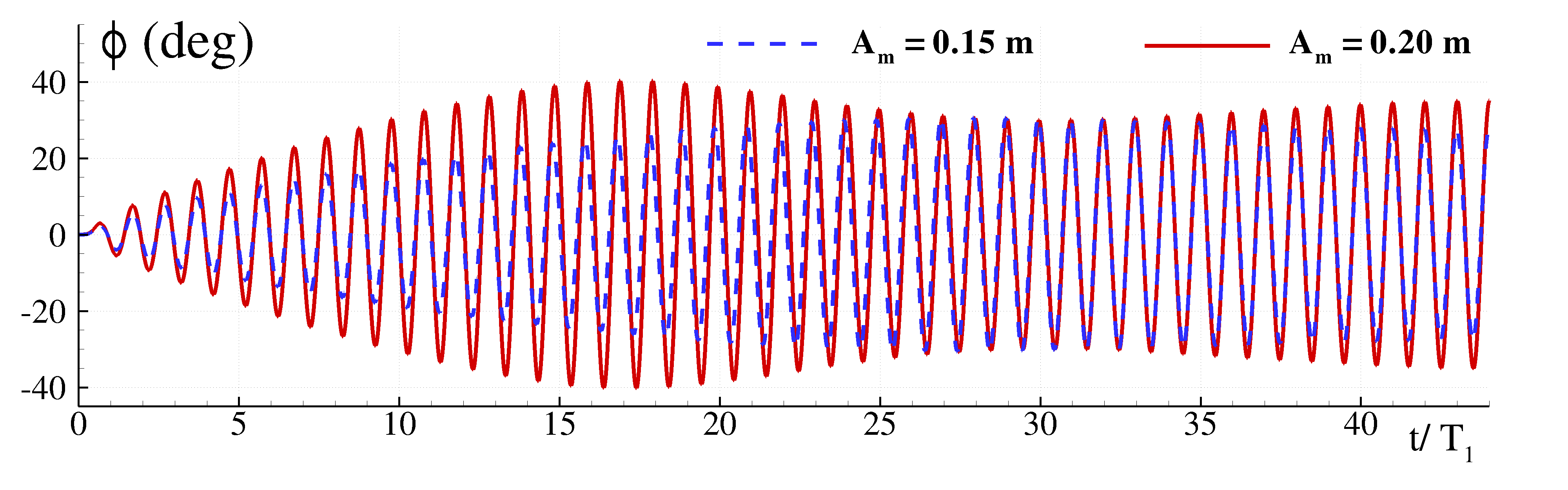}
\caption{Tank filled with water: roll angle plotted as a function of time using $A_m = 0.05$ m, $A_m = 0.1$ m (top), $A_m = 0.15$ m, $A_m = 0.2$ m (bottom), excitation frequency $\omega\,=\,\omega_1^m$.  }
\label{fig:Rollangle_water}
\end{figure}

The experimentally obtained roll angles are plotted as a function of time in Fig.  \ref{fig:Rollangle_water}.
For the first three Series, the system reaches an approximate time-periodic state
within $40$ periods; the duration of the experiment is not long enough for Series $4$ to reach such state.
%
\begin{figure}[h!]
\centering
\includegraphics[width=0.84\textwidth]{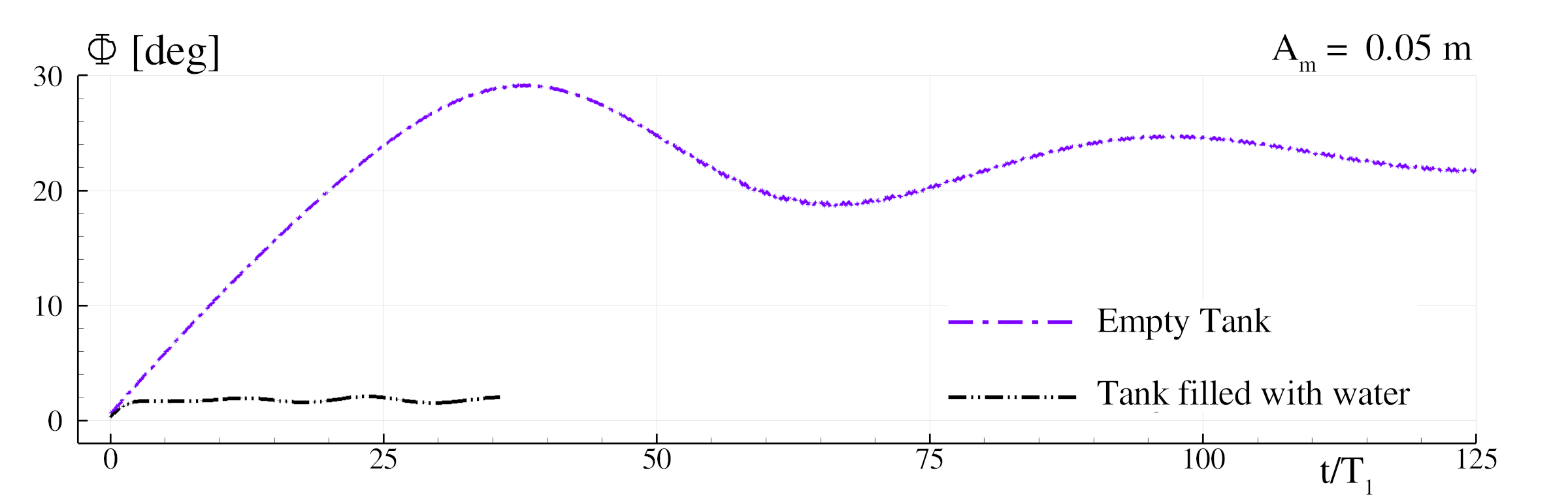}
\includegraphics[width=0.84\textwidth]{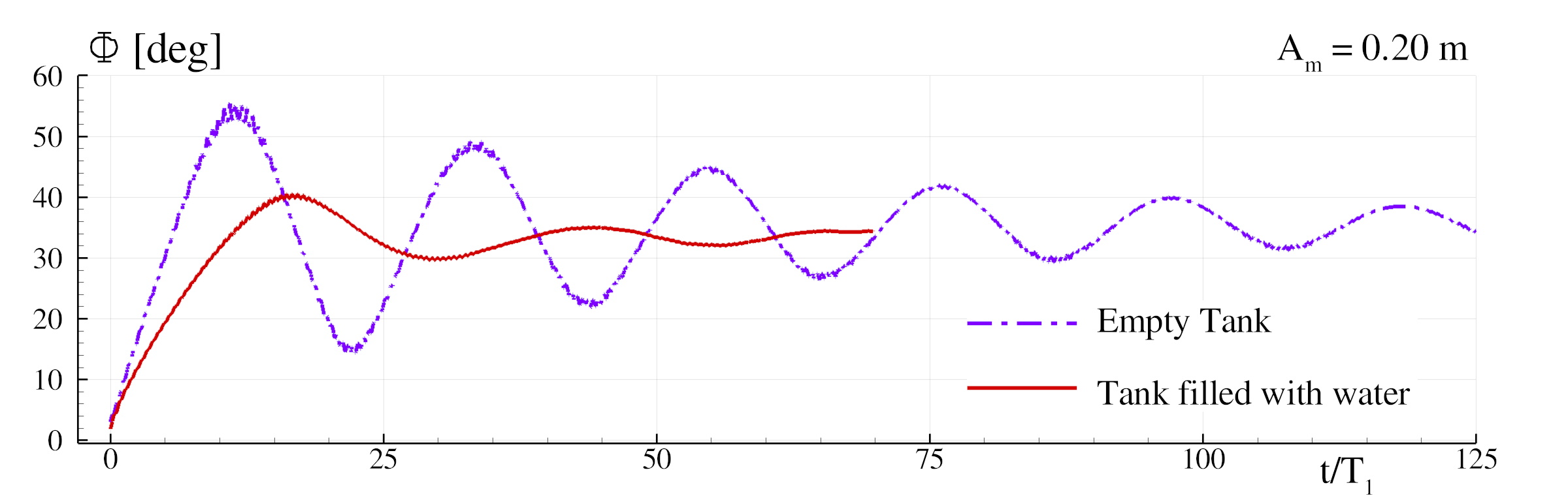}
\caption{Envelope function $\Phi$ plotted as a function of time, using excitation amplitudes
$A_m=0.05$ m (top) and $A_m=0.20$ m (bottom) for the empty tank and for the water filled tank.  }
\label{fig:Phi_water}
\vspace*{0.3cm}
\end{figure}

Fig. \ref{fig:Phi_water} depicts the envelope function $\Phi$ defined in (Part I section II.C)) as:
\begin{equation}\label{eq:phi_eff}
\Phi(t)\,=\,\frac{\pi}{2\,T}\int_{t}^{t+T}\,|\phi(s)|\,ds\,
\end{equation}
obtained with the empty tank and with the tank filled with water for both Series 1 and 4.


The top plot is obtained for the smallest excitation amplitude $A_m=0.05$ m.
In this case, the system behaves like a classical TLD, i.e. when water is present, $\Phi$ is drastically reduced. In the transient stage and for the largest excitation, $A_m=0.20$ m, the angle $\Phi$ is lower
when water is present (maximum angle $52^\circ$ with the empty tank against $40^\circ$ with water), as can be seen in the same figure.  On the other hand, at time-periodic state, the roll angle is similar for the two
different configurations (empty / filled with water).

This behavior is partially explained by the fact
that $M_{mass/tank}$ is approximately proportional
to $A_m$ (see equation (II.4) of Part I) 
while $M_{fluid/tank}$ is approximately proportional to $\sqrt{\Phi}$ (see equation (IV.27) of Part I). 
The behavior of the coupled system is further influenced by the phase lags between the different torques.
%
\subsection{Phase lags analysis of the coupled system}
\begin{figure}[t!]
\centering
\includegraphics[width=0.70\textwidth]{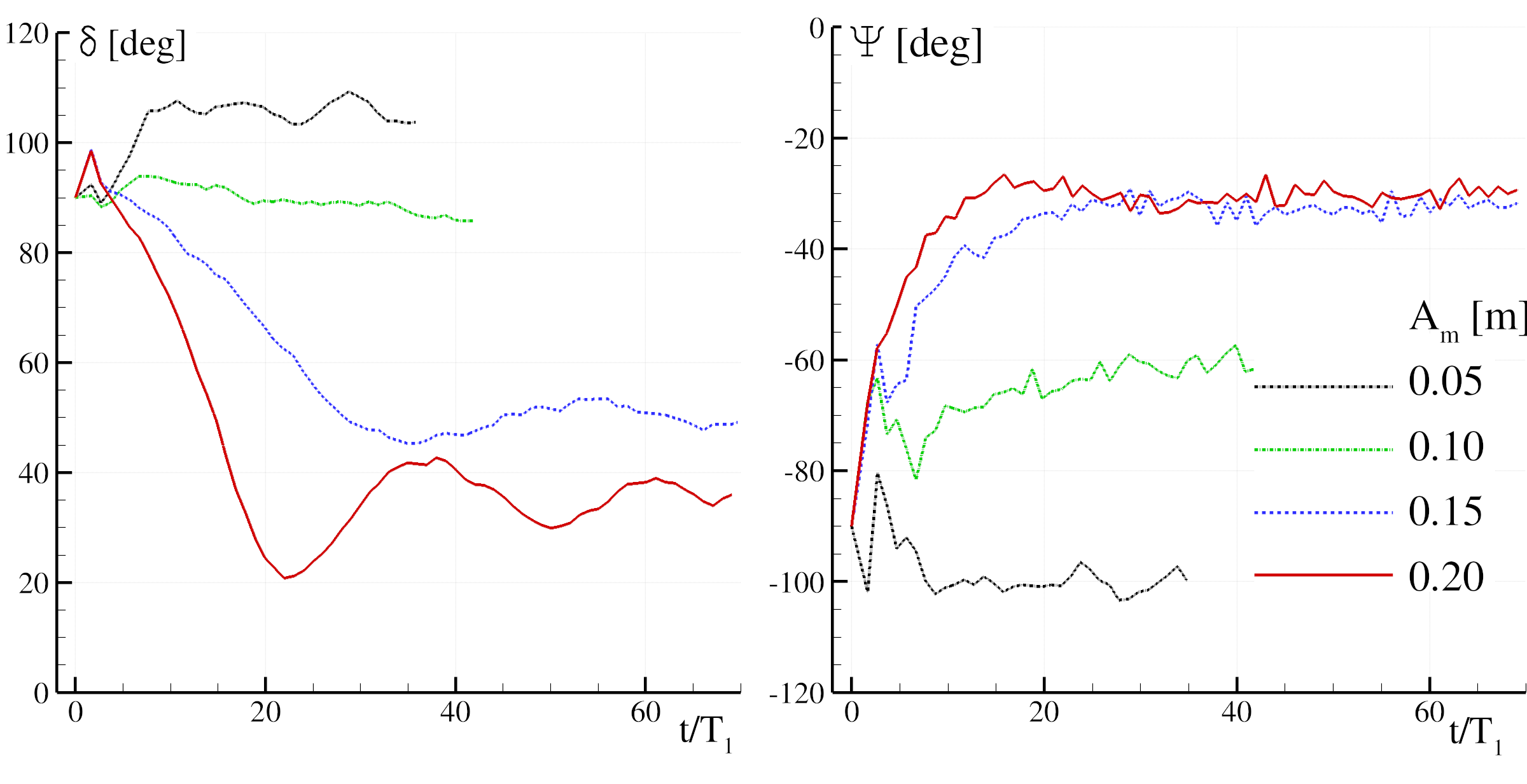}
\caption{Tank filled with water: phase lags $\delta$ and $\Psi$ plotted as a function of time using the four different excitation amplitudes $A_m$ mentioned in section \ref{sec:exp_setup}. }
\label{fig:Delta_Psi_water}
\vspace*{0.3cm}
\end{figure}

The results of the time evolution of $\delta$ (phase lag between the sliding mass motion and roll angle) and
$\Psi$ (phase lag between $M_{fluid/tank}$ and the roll angle) are shown in Fig. \ref{fig:Delta_Psi_water}.
The torque $M_{fluid/tank}$ is evaluated through (Eq. II.8 of Part I):
\begin{equation}\label{eq:M_fluid_tank}
M_{fluid/tank} \,=\,I_0 \ddot{\phi} -  g S_g \sin(\phi)  + B_{\phi} \dot{\phi} + K_{df} \sgn(\dot{\phi})\,+\,
                 m {\xi}_m  g cos(\phi) + m(2 \xi_m \dot{\xi}_m \dot{\phi} + {\xi}_m^2 \ddot{\phi})
\end{equation}
where the motions of the sliding mass $\xi_m(t)$ and of the tank $\phi(t)$ are experimentally recorded.

For the smallest value of $A_m$, the torque $M_{fluid/tank}$ (see \ref{eq:M_fluid_tank}) is approximately in quadrature with the tank motion $(\Psi\approx-90^\circ)$ during the time evolution. The
roll motion is also approximately in quadrature with the sliding mass motion ($\delta \approx 90^\circ$).
For larger $A_m$, the phase lag $\Psi$ measured at the time-periodic state moves to $\Psi=-30^\circ$ for Series 3). $\delta$ also changes accordingly with the net outcome being that $M_{fluid/tank}$ remains approximately in counter-phase with $M_{mass/tank}$ at time-periodic state. Indeed, the fluid helps the system to reduce the effect of the external excitation.

\begin{figure}[t!]
\centering
\includegraphics[width=0.70\textwidth]{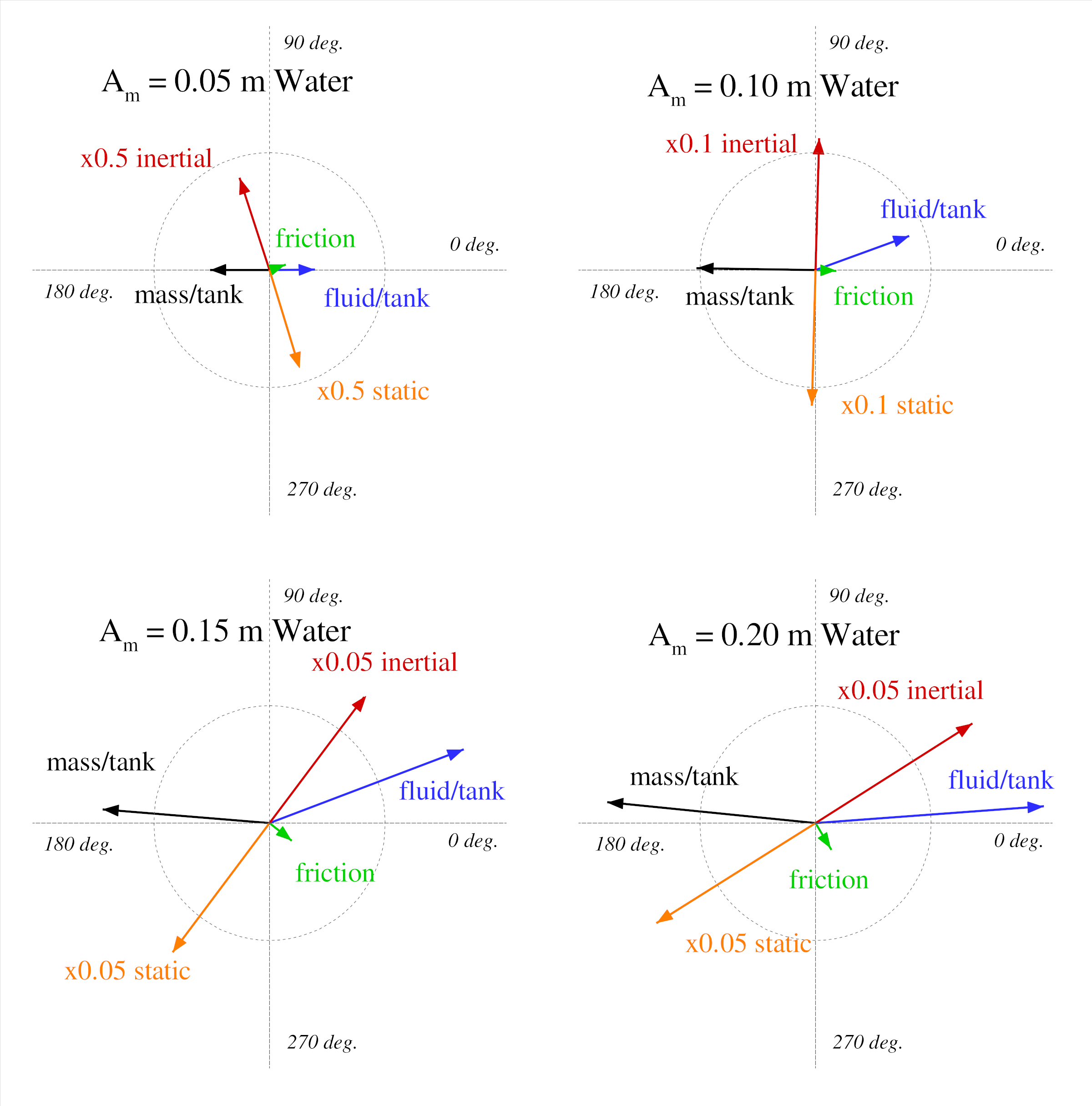}
\caption{Phasors obtained for the tank filled with water (see definition in Part I). Torque modulus and phase are retrieved as the first harmonic at time-periodic state.}
\label{fig:Phasors_water}
\end{figure}

This can be appreciated in the phasors' graphs, corresponding  to the time-periodic state, of Fig.  \ref{fig:Phasors_water}.
For the smallest value of $A_m$ (top panel), the system behaves basically like the ideal TLD system presented in section (V-A) of Part I.
For the largest value of $A_m$ (bottom panel), the behavior is completely different from what expected in a ideal TLD.

In Part I, the dissipation properties of the Pendulum-TLD excited with large values of $A_m$ was discussed and the analogy of the present system with a hybrid mass liquid damper was established. With such an analogy it is important that the damper does not return energy to the external system.
As it shown in Part I, this is not always the case with the empty tank.
This implies that it is necessary to analyze energy transfers during the process in order to establish the practical interest of the present system.

\subsection{Energy transfer from tank to fluid}
\begin{figure}[ht!]
\centering
\includegraphics[width=0.80\textwidth]{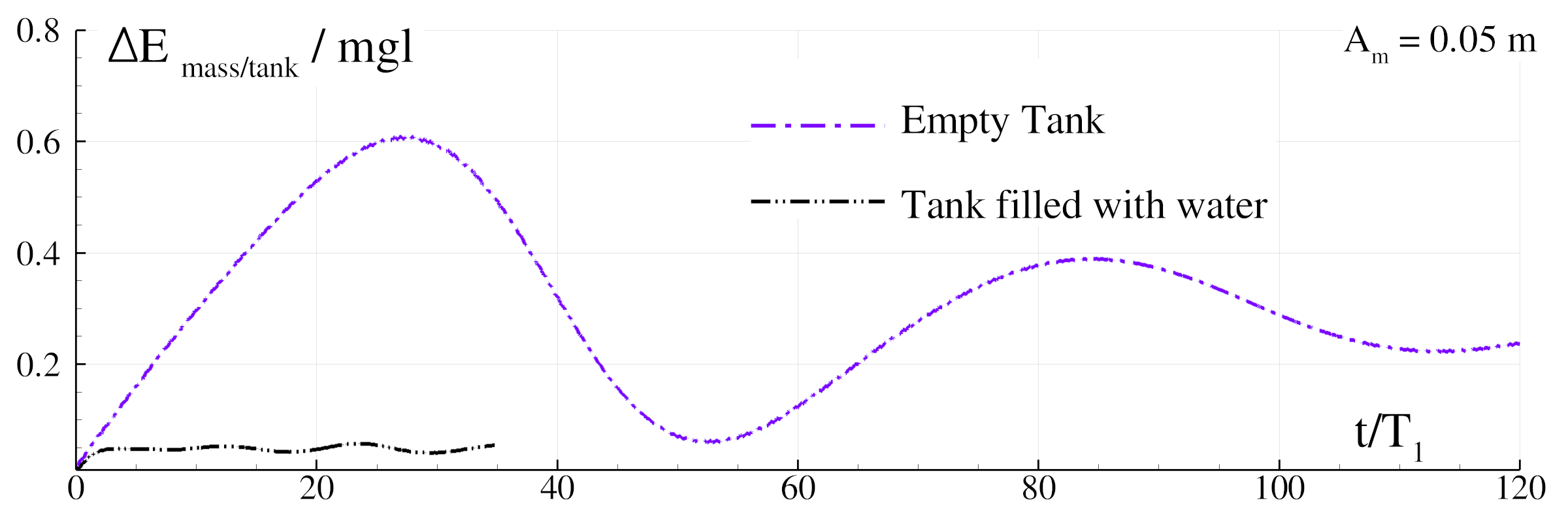}
\vskip 0.5cm
\includegraphics[width=0.80\textwidth]{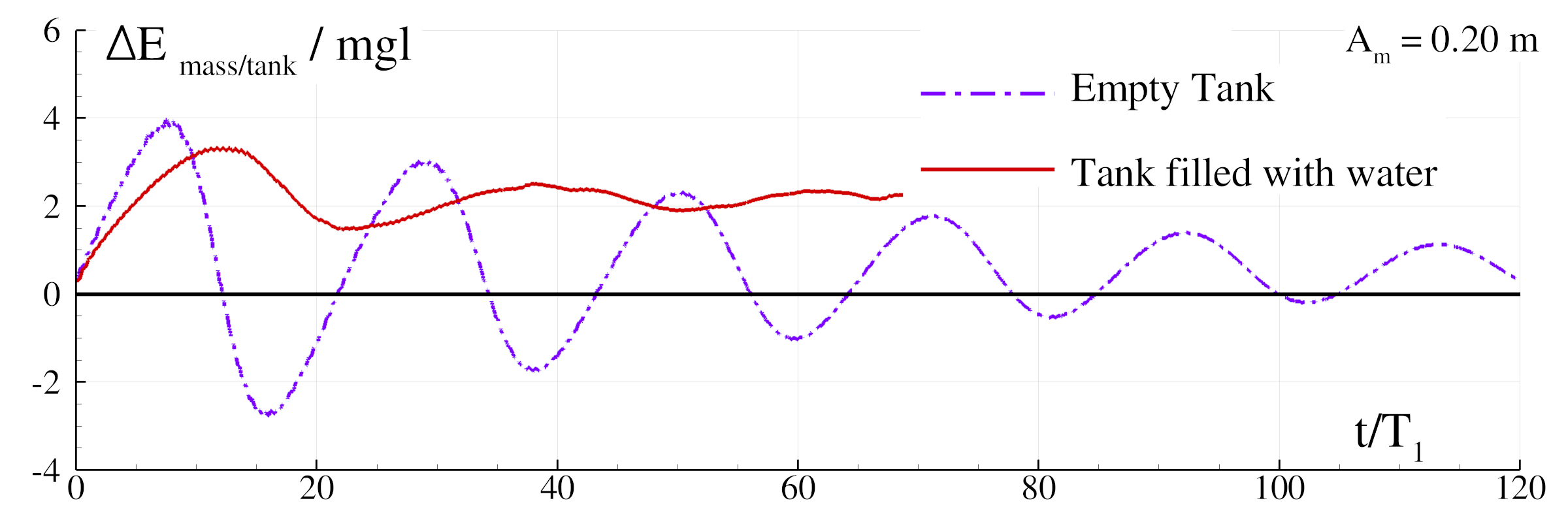}
\caption{$\Delta E_{mass/tank}$ plotted as a function of time, using excitation amplitudes
$A_m=0.05$ m (top) and $A_m=0.20$ m (bottom) for the empty tank and for the tank filled with water.}
\label{fig:Delta_E_mass_tank_water}
\end{figure}

The time history of the energy transfer between the sliding mass and the tank in one cycle, $\Delta E_{mass/tank}$, can be calculated with equation:
\begin{equation}\label{eq:deltaEmasstank}
\Delta\,E_{mass/tank}\,=\,\int_{t}^{t+T}\,M_{mass/tank}(s)\,\dot{\phi}\,ds,
\end{equation}
This variable is depicted in Fig. \ref{fig:Delta_E_mass_tank_water} for the extreme sliding mass motion amplitudes
(the reader is referred to section II of Part I for all the torque and energy term definitions and sign conventions).
%

For Series 1, when the tank is empty, $\Delta E_{mass/tank}$ is always positive
and the value largely oscillates before ending in a time-periodic state. When water is in the tank, $\Delta E_{mass/tank}$ is drastically reduced and so is the roll angle amplitude. The low level of $\Delta E_{mass/tank}$ is continuously transferred to the gentle sloshing flow regime.

For Series 4, when the tank is empty, $\Delta E_{mass/tank}$ periodically changes in sign;
in other words, the sliding mass exerts work on the tank during some parts
of the process but also receives work from the tank in others. This is a major
concern, assuming that the goal is to retrieve energy from the exciting system.
With this amplitude and the empty tank, the sign only stabilizes as positive after many periods.
On the other hand, in presence of water, $\Delta E_{mass/tank}$ remains positive;
the sliding mass  always makes a positive and intense work on the tank
which is then transferred to the fluid through $-\Delta E_{fluid/tank}$.  At time-periodic state
this work is fully dissipated by the fluid. This dissipation can be quantified, as discussed hereafter.
%
%
\subsection{Wave breaking and energy dissipation}
\begin{figure}[ht!]
\centering
\includegraphics[width=0.80\textwidth]{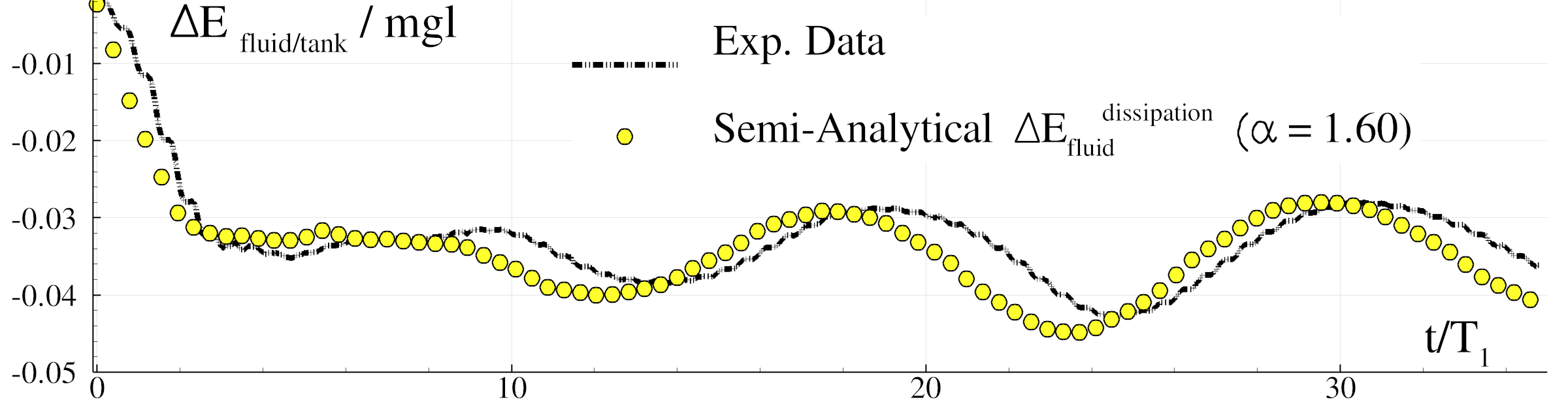}
\vskip 0.3cm
\includegraphics[width=0.80\textwidth]{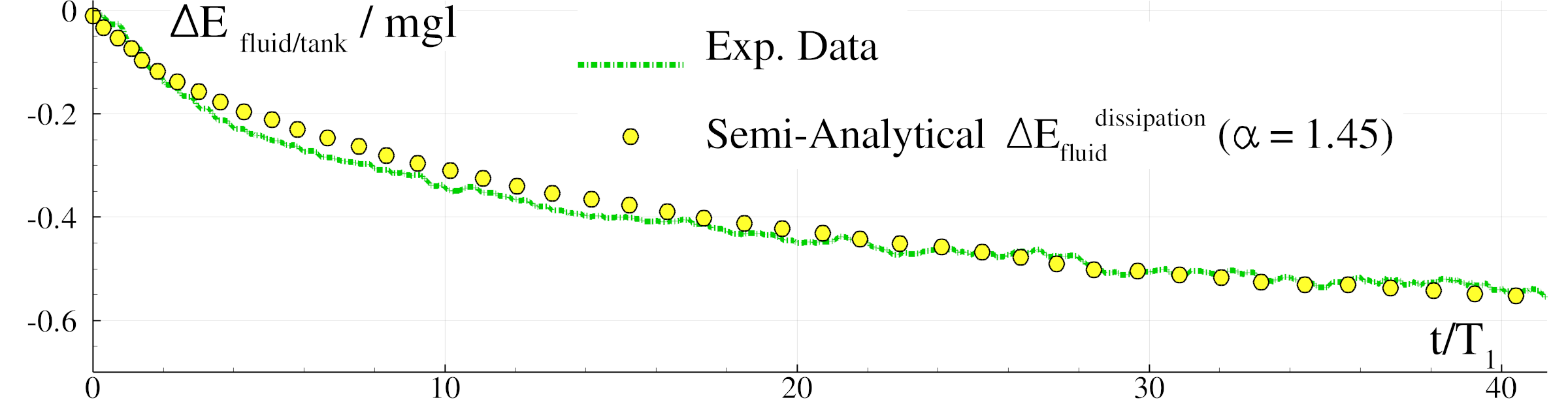}
\vskip 0.3cm
\includegraphics[width=0.80\textwidth]{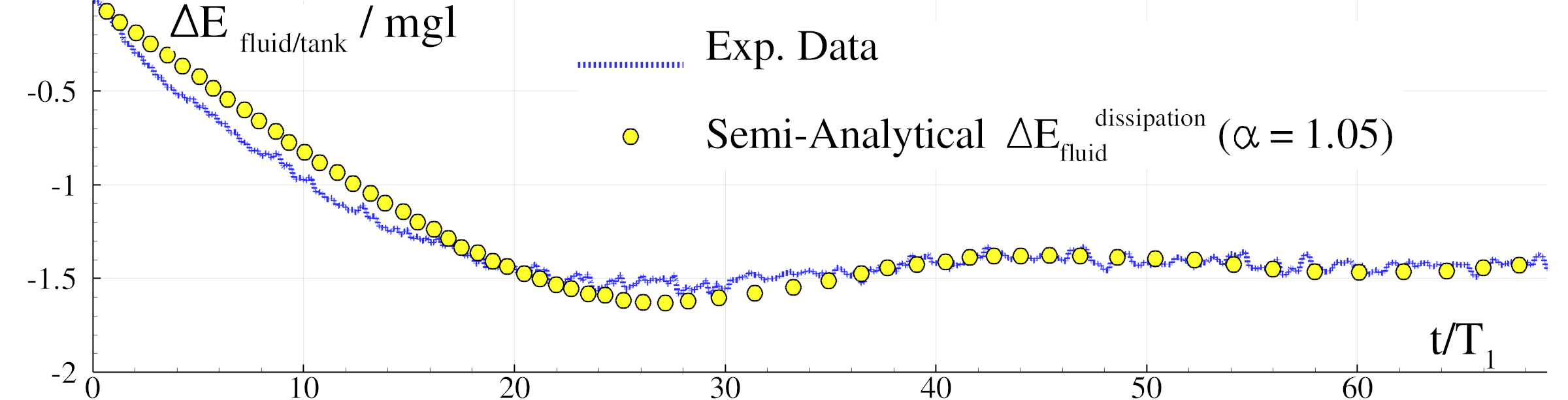}
\vskip 0.3cm
\includegraphics[width=0.80\textwidth]{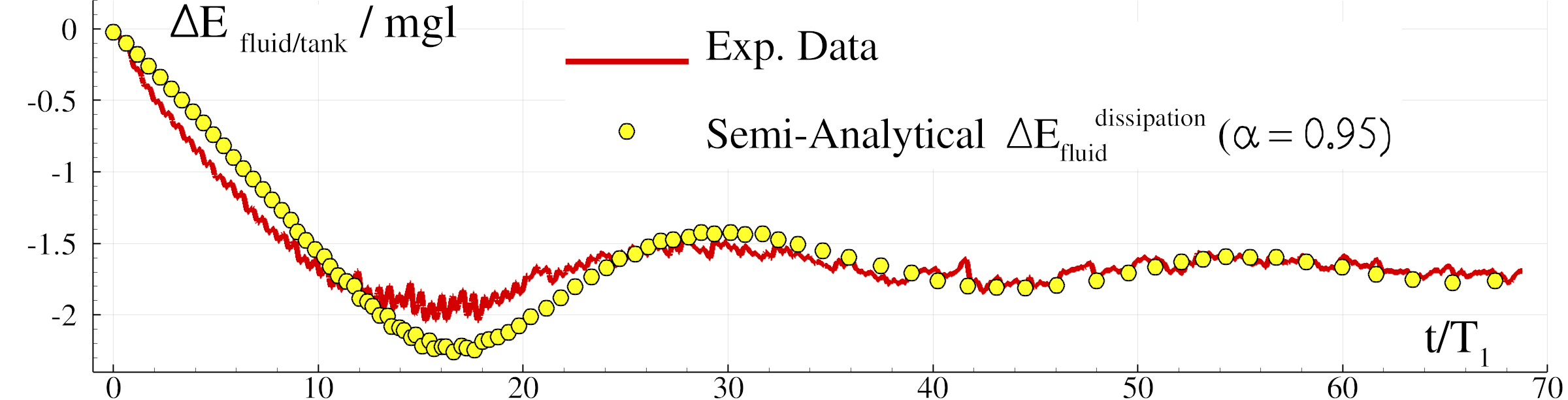}
\caption{
Energy transfer $\Delta E_{fluid/tank}$ plotted as a function of time, using excitation amplitudes
$A_m=$ 0.05, 0.10, 0.15 and 0.20 m (from top to bottom).
}\label{fig:Delta_E_fluid_tank_water}
\end{figure}

The energy transfer between the fluid and the tank, $\Delta E_{fluid/tank}$, is obtained as:
\begin{equation}\label{eq:E_fluid_tank}
\Delta\,E_{fluid/tank}\,:=\,\int_{t}^{t+T}\,M_{fluid/tank}(s)\,\dot{\phi}\,ds,
\end{equation}
where $M_{fluid/tank}(s)$ is computed from experiments using eq. (\ref{eq:M_fluid_tank}). Fig. \ref{fig:Delta_E_fluid_tank_water} plots $\Delta E_{fluid/tank}$ as a function of time for the four Series. This term comprises the fluid mechanical energy variation and a dissipation term (see also Part I):
\begin{equation}\label{eq:P_fluid_dissipation}
\Delta\,E_{fluid/tank}\,=\,-[E_{fluid}^{mech}]_t^{t+T}\,+\,\Delta\,E_{fluid}^{dissipation}
\end{equation}
%
%
On these plots a dissipated energy per cycle, $\Delta E_{fluid}^{dissipation}$,
time history plot is also included computing it with the following semi-analytical expression (see Part I section IV-C):
\begin{equation}\label{Energy_SPH_Verhagen}
\Delta E_{fluid}^{dissipation}(t)\,=\,-\alpha\,\left[\,4\,m_{liquid}\,g\,h\,\Phi^{\frac{3}{2}}(t)\,\right]\,,
\end{equation}
in which $\Phi(t)$ is the recorded envelope (see eq. (\ref{eq:phi_eff})) and $\alpha$ is a constant coefficient $\alpha$
tuned in order to match the curves on the final part of the time histories.
The range of $\alpha$ adopted is in agreement with what was obtained in the
numerical analysis for forced motion performed in section IV-D of Part I.
Indeed, for the smaller excitation amplitude cases, $A_m =$ 0.05 m,
the coefficient $\alpha$ is close to the theoretical value $1.68$ found in Part I.
By increasing the amplitude, $\alpha$ is progressively reduced as predicted by the SPH model discussed
in Part I.

Even if these ``semi-analytical'' curves may be a crude approximation, the fine match with the experimental
data indicates:
\begin{enumerate}
 \item the predicted values for the dissipation in time-periodic state for water are within the range found in Part I.
 \item that for moderate and large amplitude cases, the large increases in dissipation are linked to the breaking wave phenomena, that dominate the flow dynamics.
 \item the component $[E_{fluid}^{mech}]_t^{t+T}$ seems to be much smaller than $\Delta\,E_{fluid}^{dissipation}$ along the whole process, except for the very early stage. This is also confirmed by other results shown in section \ref{Summarydissipation}.
\end{enumerate}
%
%

It is observed that a perfect time-periodic state is not reached for Series 1 and a sub-harmonic behavior appears.
However, the amplitudes of the sub-harmonics are small and the phenomenon can be ignored in the present analysis.

Using the results of the previous subsection, it can be stated that the dissipation associated with intense breaking wave phenomena stabilizes the sign of the energy $\Delta E_{mass/tank}$ for Series 2, 3 and 4.
For Series 1 the low level of energy involved only induces the development of wave trains inside the tank.

%
\subsection{Summary of results for the water case}
To summarize the results of this section, the value of the main quantities at time-periodic state
are reported in table \ref{tab:water}.
It can be noted that the energy transfers $\Delta E_{fluid/tank}$ and $\Delta E_{mass/tank}$  for the case $A_m=0.20$ m are almost $50$ times higher than the one with $A_m=0.05$ m, while the external work ratio $\Delta E_{mass/tank}^{water}/\Delta E_{mass/tank}^{empty}$ increases by a factor of 25.

\begin{figure}[ht!]
\centering
\includegraphics[width=0.90\textwidth]{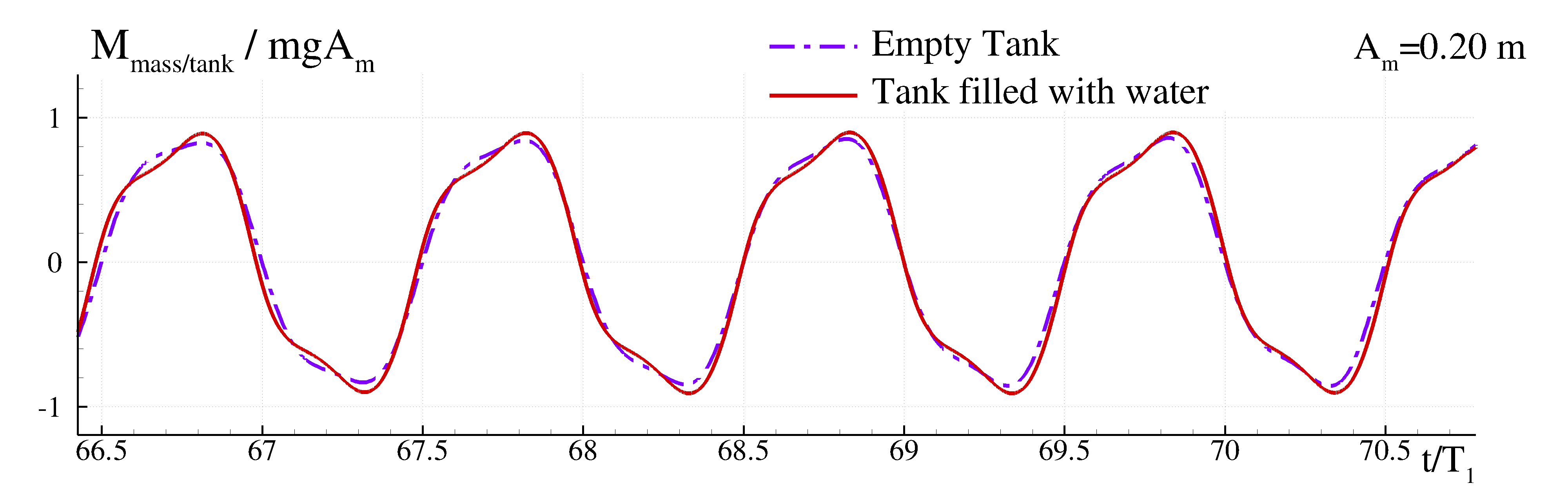}
\caption{Torque $M_{mass/tank}$ plotted as a function of time, using excitation amplitude
$A_m=0.20$ m for the empty tank and for the water filled tank.}
\label{fig:Mmasstank_water}
\end{figure}

Finally, in Fig. \ref{fig:Mmasstank_water} the torque $M_{mass/tank}$ is plotted as a function of time
for the excitation amplitude $A_m=0.20$ m for both the empty tank and for the tank filled with water.
This plot confirms that even when the water is present, $M_{mass/tank}$ remains in phase opposition
with the sliding mass motion. It also confirms that the torque amplitude is practically unaffected by the presence
of water (\emph{i.e.} the differences induced by the roll motion $\phi(t)$ on the torque $M_{mass/tank}$ are negligible).
%
\begin{table}[ht!]
\begin{center}
    \begin{tabular}{|l||c|c|c|c|}
        \hline
        $A_m$  [m]                  & 0.05   & 0.10  & 0.15  & 0.20    \\ \hline \hline
        $\Phi$ [deg]             & 1.7    & 12    & 28    & 33    \\ \hline
        $\delta$ [deg]           & 105    & 85    & 50    & 35    \\ \hline
        $\Psi$ [deg]             & -100   & -60   & -33   & -30   \\ \hline
        $\Delta E_{mass/tank}/mgl$  & 0.05   & 0.65  & 1.8   & 2.2   \\ \hline
        $\Delta E_{fluid/tank}/mgl$ & -0.035 & -0.52 & -1.42 & -1.68 \\ \hline
        $\Delta E_{friction}/mgl$   & -0.015 & -0.13 & -0.38 & -0.52 \\ \hline \hline
        $\Phi^{water}/\Phi^{empty}$  & 0.085  & 0.43  & 0.90  & 0.97 \\ \hline
        $\Delta E_{mass/tank}^{water}/\Delta E_{mass/tank}^{empty}$
                                    & 0.17   & 1.55  & 3.6   & 4.07  \\ \hline
    \end{tabular}
\caption{\label{tab:water} Tank filled with water: values of the main quantities reached at time-periodic state for the excitation amplitudes: $A_m$= 0.05, 0.10, 0.15 and 0.20 m.}
\end{center}
\vskip 0.2 cm
\end{table}
%
\section{Effects of the liquid adopted: viscosity and density.} \label{ss:fluid_viscosity}

In the performed experiments, the influence of the viscosity of the fluid is explored. Table \ref{fluid_matrix} records the physical properties of the three liquids used: water, sunflower oil, and glycerine. The total mass of fluid in the tank, the surface tension coefficient, and the density, all differ little among the three liquids. But the three values of the kinematic viscosity vary such that the oil is fifty times more viscous than water, and the glycerine is 740 times more viscous than water.

\begin{table}[ht!]
\begin{center}
\vspace*{0.2cm}
\begin{tabular}{|c||c|c|c|c|}
\hline
                     & $\nu (m^2 \cdot s^{-1})$  & $\sigma (mN/m)$ & $\rho (kg \cdot m^{-3})$ & $m_{liquid} (Kg)$\\
\hline Glycerine      & $7.4\cdot 10^{-4}$        & 64 & 1261  & 6.474     \\[2pt] 
\hline Sunflower Oil & $5\cdot 10^{-5}$          & 33 & 900   & 4.620     \\[2pt] 
\hline Water         & $10^{-6}$                 & 72 & 998   & 5.123     \\[2pt] 
\hline
\end{tabular}
\vskip 0.3cm
{\footnotesize \caption{\label{fluid_matrix} kinematic viscosity, surface tension, density and mass inserted inside the tank for the studied fluids.}}
\end{center}
\end{table}

In Table \ref{tab:Re_sloshing} the different Reynolds numbers, defined as $Re=\sqrt{gh}\,A_m/\nu$,
and Weber number, defined as $We =\rho\,gh\,A_m/\sigma$, are reported, assuming
$\sqrt{gh}$ is the characteristic flow velocity in shallow water condition.
\begin{table}[ht!]
\begin{center}
\vspace*{0.2cm}
\begin{tabular}{|c|c||c|c|c||c|c|c|c|}
\hline
                     & $A_m$ (m) & $Re^{water}$ & $Re^{oil}$ & $Re^{glycerine}$
                             & $We^{water}$ & $We^{oil}$ & $We^{glycerine}$
                             \\ \hline
\hline Series $1$  & $0.05$  &  47500 & 950  & 64  & 625   &  1230  &   890 \\[2pt]
\hline Series $2$  & $0.10$  &  95000 & 1900 & 128 & 1250  &  2460  &  1780 \\[2pt]
\hline Series $3$  & $0.15$  & 142500 & 2850 & 193 & 1880  &  3690  &  2670 \\[2pt]
\hline Series $4$  & $0.20$  & 190000 & 3800 & 256 & 2500  &  4920  &  3560\\[2pt]
\hline
\end{tabular}
\vskip 0.3cm
{\footnotesize \caption{\label{tab:Re_sloshing} Reynolds and Weber numbers for the sloshing flows induced in the oscillating tank when using water, sunflower oil and glycerine inside the tank.}}
\end{center}
\end{table}

As shown hereinafter, when using sunflower oil and glycerine, the roll angle obtained is generally larger than the one recorded with water. As a consequence and for safety reasons, the tests for Series 3 and Series 4 are stopped during the transient, before a time-periodic state could be reached.
%
\subsection{Series 1}
\begin{figure}[ht!]
\centering
\includegraphics[width=0.31\textwidth]{Water_A50_A}
\includegraphics[width=0.31\textwidth]{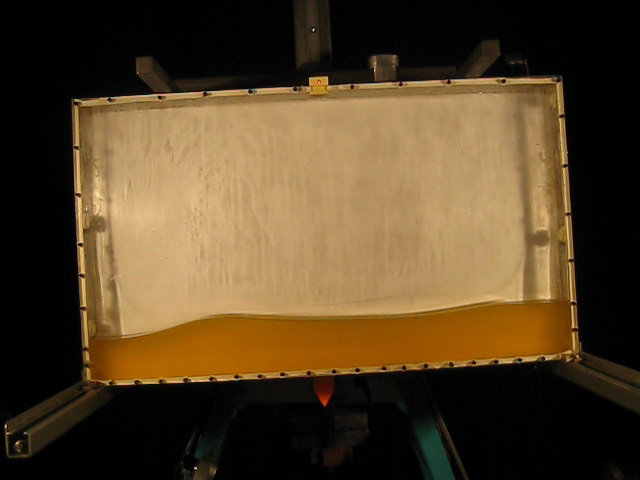}
\includegraphics[width=0.31\textwidth]{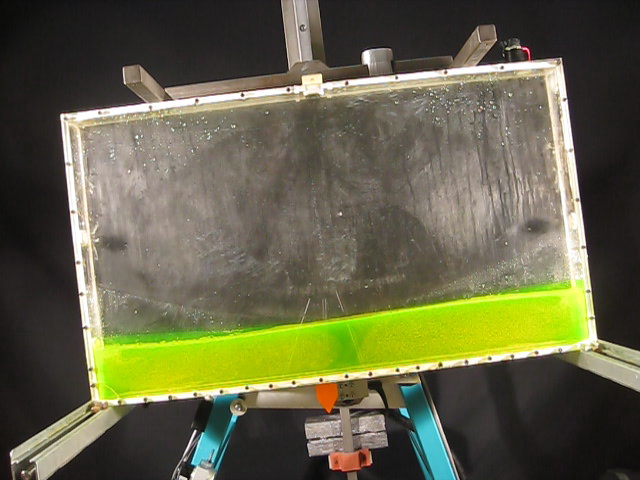}
\includegraphics[width=0.31\textwidth]{Water_A50_B}
\includegraphics[width=0.31\textwidth]{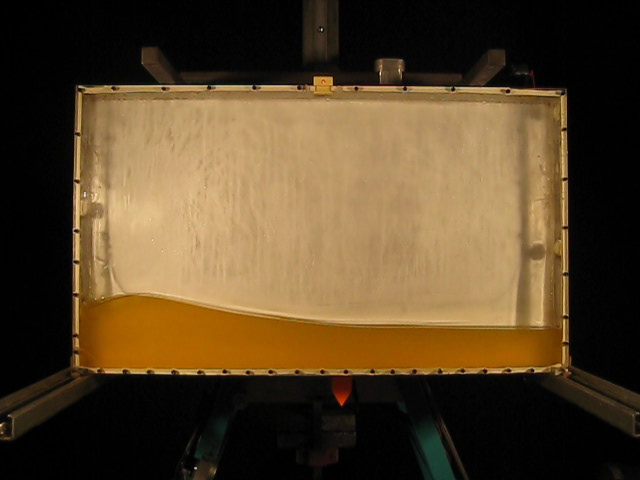}
\includegraphics[width=0.31\textwidth]{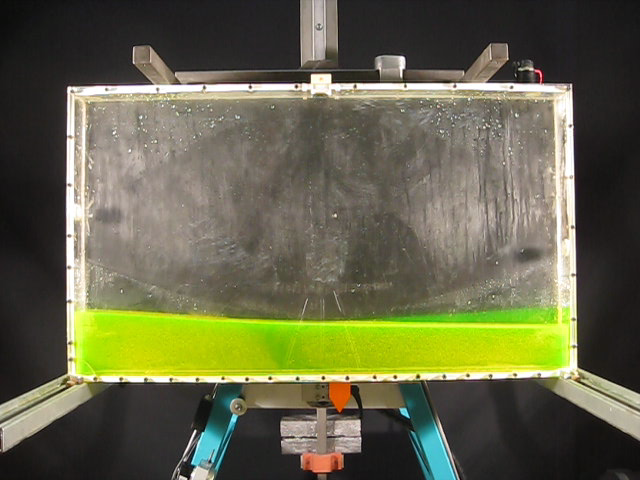}
\caption{Sloshing wave in tank:  filled with water, oil and glycerine $A_m = 0.05 m$. See supplementary materials from
[URL will be inserted by AIP].
} \label{fig:EvolSlosh_A50}
\end{figure}

Two pictures, representative of the flow evolution at time-periodic state for Series 1, are shown in Fig. \ref{fig:EvolSlosh_A50} for all three fluids. The wave train obtained with water has already been discussed.
For the oil case, the viscosity damps out the wave train and wave steepness is
drastically reduced. When using glycerine, the free surface remains almost flat.

\begin{figure}[ht!]
\centering
\includegraphics[width=0.400\textwidth]{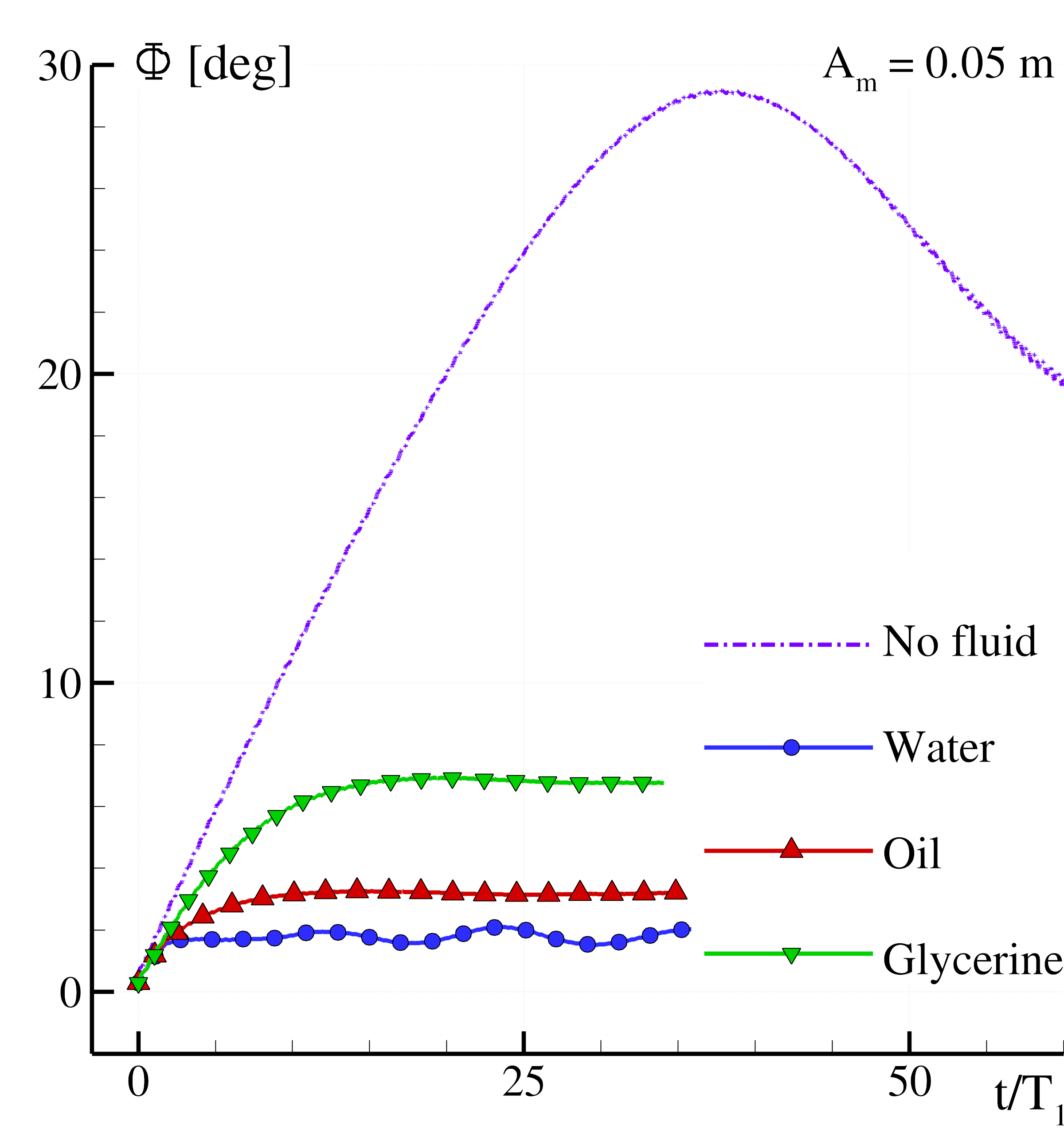}
\includegraphics[width=0.400\textwidth]{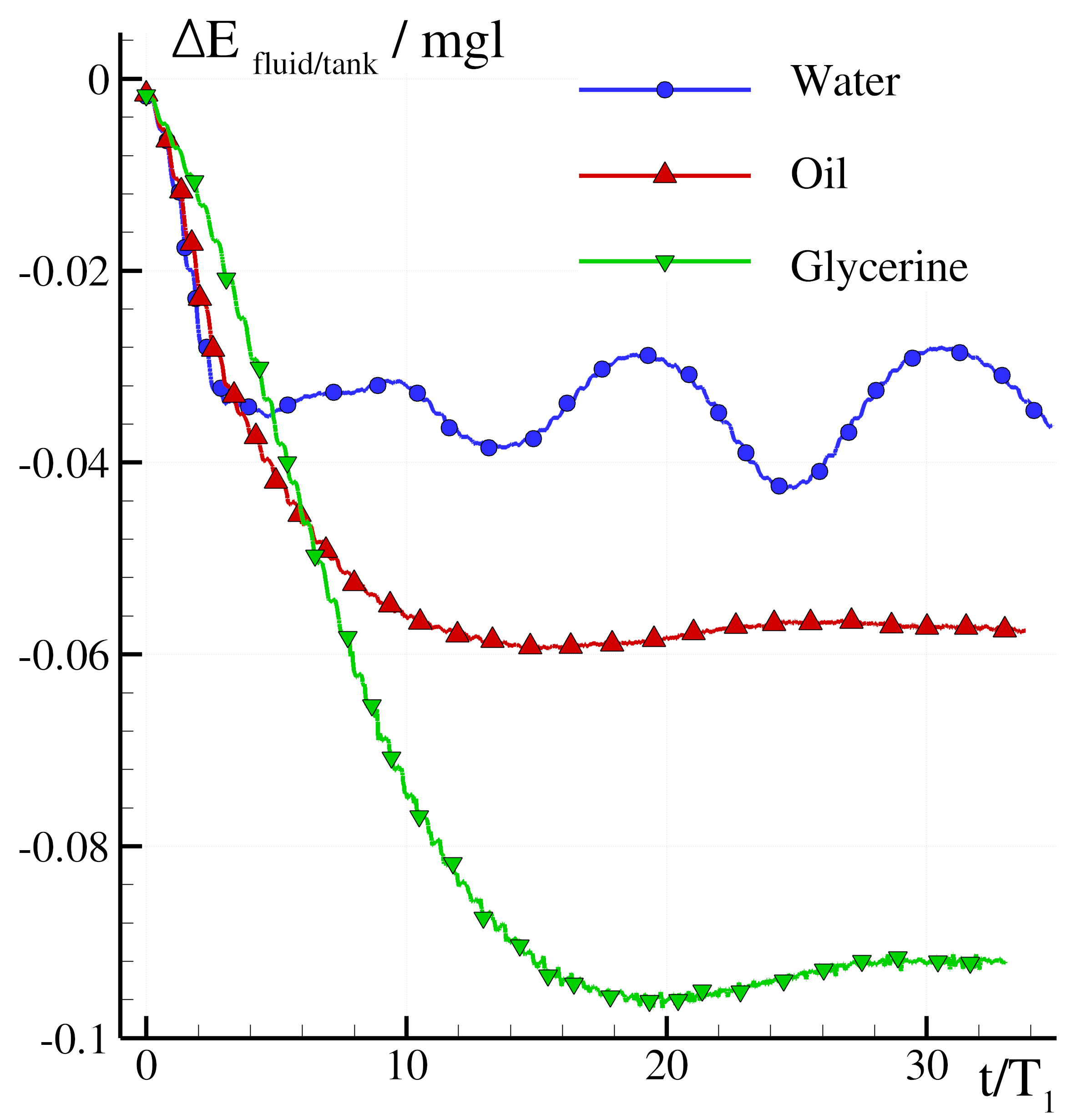}
\caption{Envelope function $\Phi$ (left) and energy transfer $\Delta E_{fluid/tank}$ (right) plotted as a function of time, with an excitation amplitude $A_m=0.05$ m,
using water, sunflower oil and glycerine inside the tank.
}
\label{fig:Delta_E_fluid_tank_oil_glycerine}
\end{figure}

The left plot of Fig. \ref{fig:Delta_E_fluid_tank_oil_glycerine} shows the
roll angle amplitude $\Phi$ as a function of time for $A_m = 0.05$ m with water, sunflower oil and glycerine.
An almost time-periodic state is achieved for all cases within about 20 periods.
The time-periodic state angles are $2$, $3$ and $7^\circ$ for water, oil and glycerine, respectively.
Therefore, the least viscous fluid (water) leads to the lowest roll angles,  and the most viscous (glycerine)
to the largest ones.
The time-periodic state angle obtained with glycerine is more than twice the one obtained with oil and more than three times that of water. These results are in agreement with those reported in \citet{Pirner2007} who found a similar behavior with methanol, water and glycerine.
This means that, if the dynamical system is considered as a TLD, the use of water allows
for the best reduction of the roll angle for a given harmonic torque at $\omega_1$.

The right plot of Fig. \ref{fig:Delta_E_fluid_tank_oil_glycerine} depicts the
energy transfer $\Delta E_{fluid/tank}$ as a function of time.
This plot shows that using water, the response of the energy transfer is accelerated, whereas using glycerine $\Delta E_{fluid/tank}$ takes more time to stabilize.
\begin{table}[ht!]
\begin{center}
    \begin{tabular}{|l||c|c|c|}
        \hline
        $Liquid$                     & water  & oil   & glycerine   \\ \hline \hline
        $\Phi$ [degree]              & 1.7    & 3.2   & 6.8         \\ \hline
        $\delta$ [degree]            & 105    & 110   & 125         \\ \hline
        $\Psi$ [degree]              & -100   & -103  & -127        \\ \hline
        $\Phi^{liquid}/\Phi^{empty}$ & 0.085  & 0.16  & 0.34        \\ \hline
        $\Delta E_{mass/tank}^{liquid}/\Delta E_{mass/tank}^{empty}$
                                    & 0.17   & 0.29  & 0.52         \\ \hline
    \end{tabular}
\caption{\label{tab:oil_glycerine} Tank filled with water, oil, glycerine: values of the main quantities reached at time-periodic state for the excitation amplitude $A_m$= 0.05 m.}
\end{center}
\end{table}

\begin{figure}[ht!]
\centering
\includegraphics[width=0.90\textwidth]{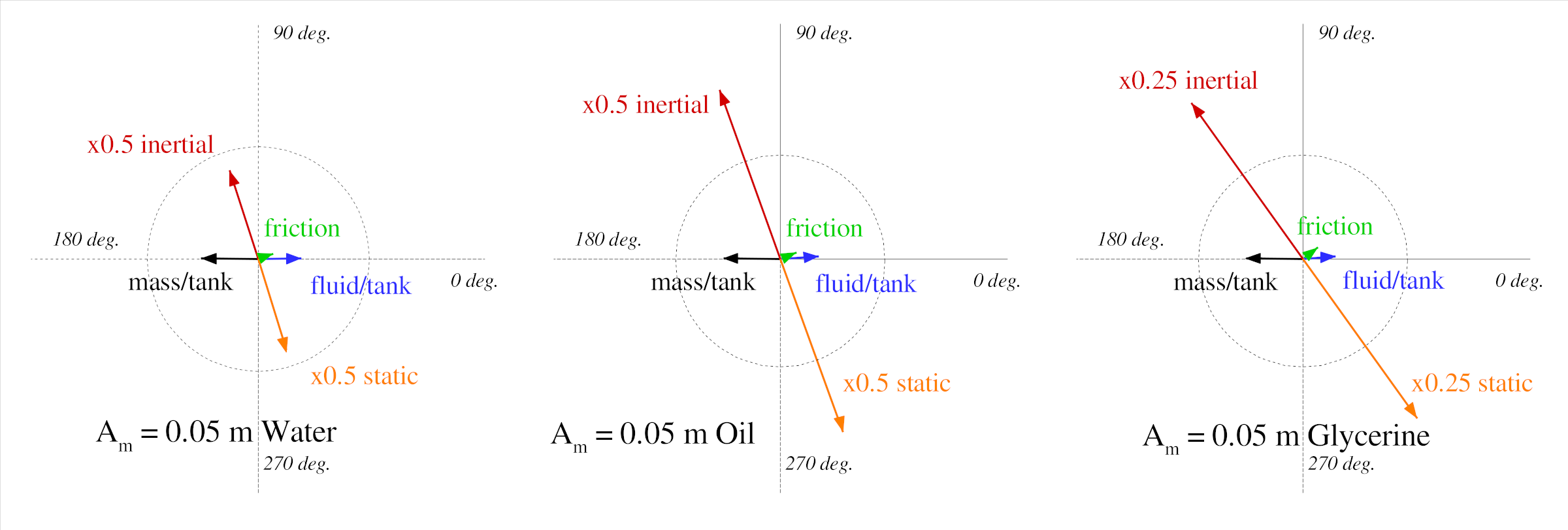}
\caption{Phasors obtained for the tank filled with water $A_m = 0.05 m$. Torque modulus and phases are retrieved as the first harmonic at time-periodic state.}
\label{fig:Phasors_fluidsAm05}
\end{figure}

The main quantities at time-periodic state using the three different liquids are
reported in table \ref{tab:oil_glycerine}.
From these measurements, it is observed that by increasing the viscosity, the moduli of $\delta$ (positive) and $\Psi$ (negative) also increase, as shown with the phasors of the various torques  plotted in Fig. \ref{fig:Phasors_fluidsAm05}. The phasor plane obtained with water has the closest
behavior to an ideal TLD such as the one discussed in Part I. This justifies why using water is a more efficient option for the reduction of the roll angle than using other liquids.
%
\subsection{Series 2} \label{sss:a01}
\begin{figure}[ht!]
\centering
\includegraphics[width=0.31\textwidth]{Water_A100_A}
\includegraphics[width=0.31\textwidth]{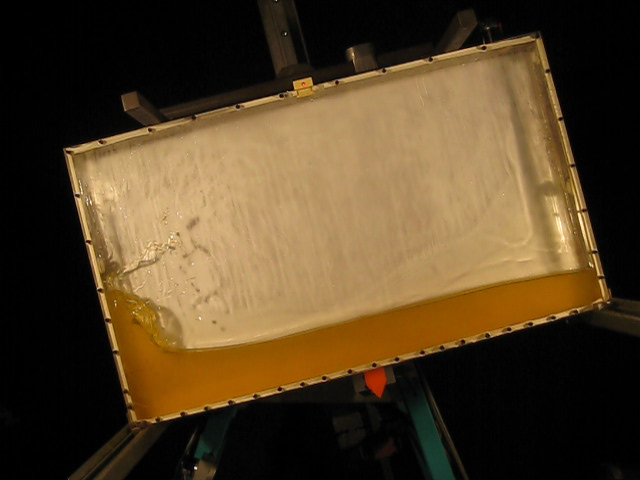}
\includegraphics[width=0.31\textwidth]{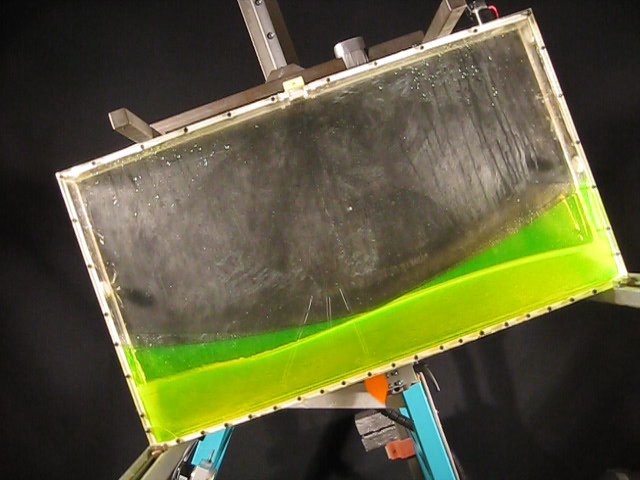}
\includegraphics[width=0.31\textwidth]{Water_A100_B}
\includegraphics[width=0.31\textwidth]{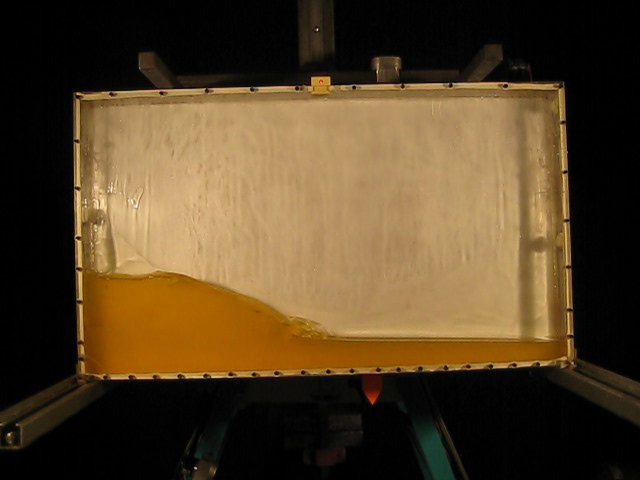}
\includegraphics[width=0.31\textwidth]{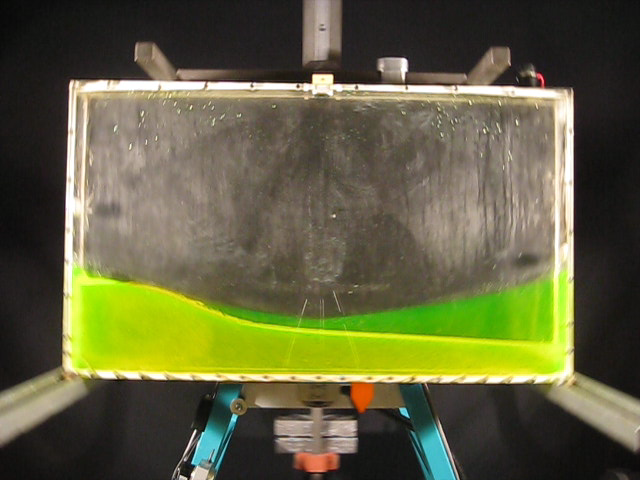}
\caption{$A_m = 0.1 m$; sloshing wave in tank:  filled with water, oil and glycerine. See supplementary materials 
from [URL will be inserted by AIP].
}
\label{fig:EvolSlosh_A100}
\end{figure}

Typical pictures of the sloshing flow at time-periodic state for Series 2
are shown in Fig. \ref{fig:EvolSlosh_A100} for all three fluids.
For Series 2 and with water, an energetic plunger develops in the middle of the tank.
When oil is present, a spilling breaking phenomena develops.
When using glycerine, the high viscosity inhibits the formation of steep waves.

\begin{figure}[ht!]
\centering
\includegraphics[width=0.400\textwidth]{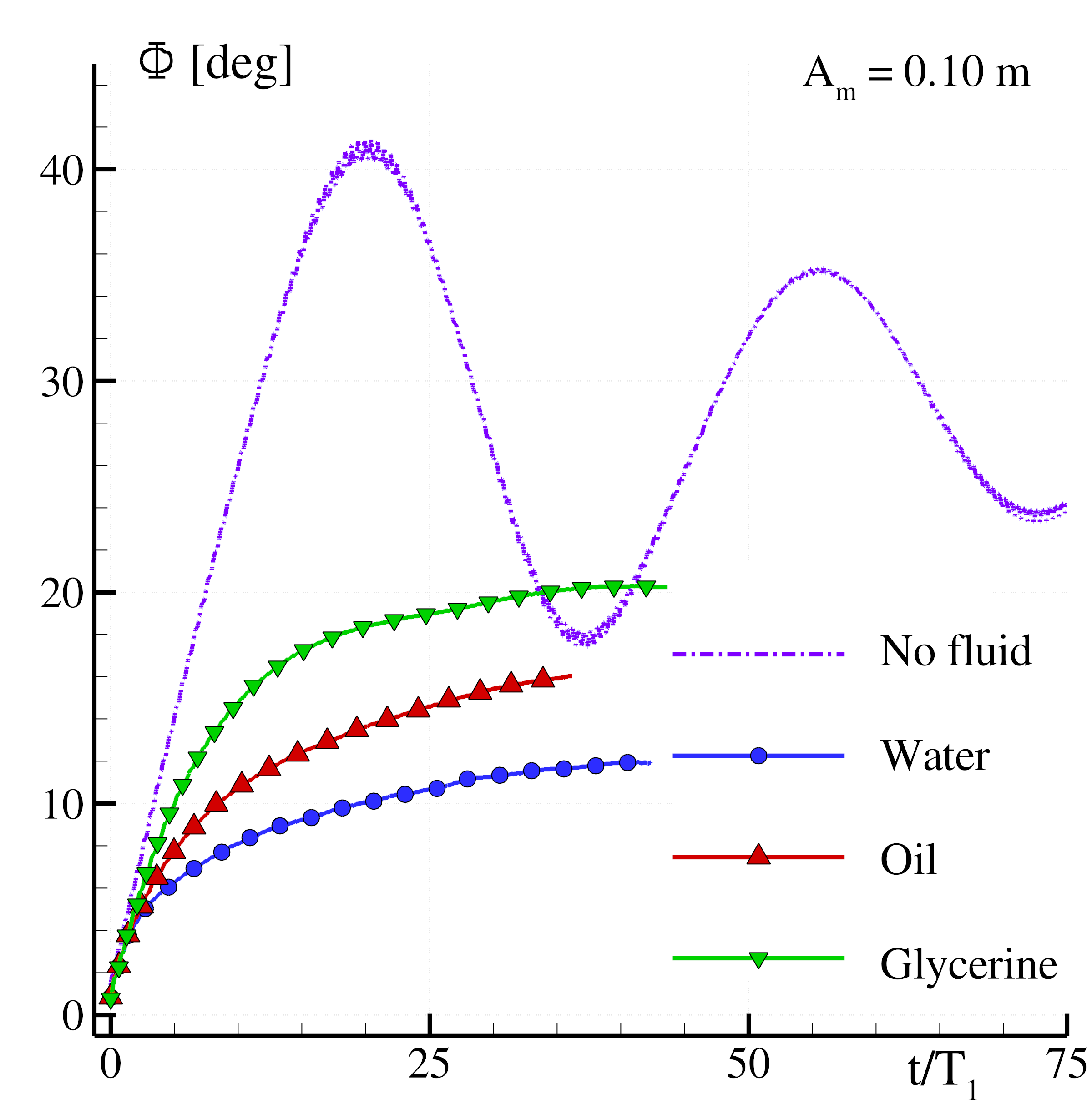}
\includegraphics[width=0.400\textwidth]{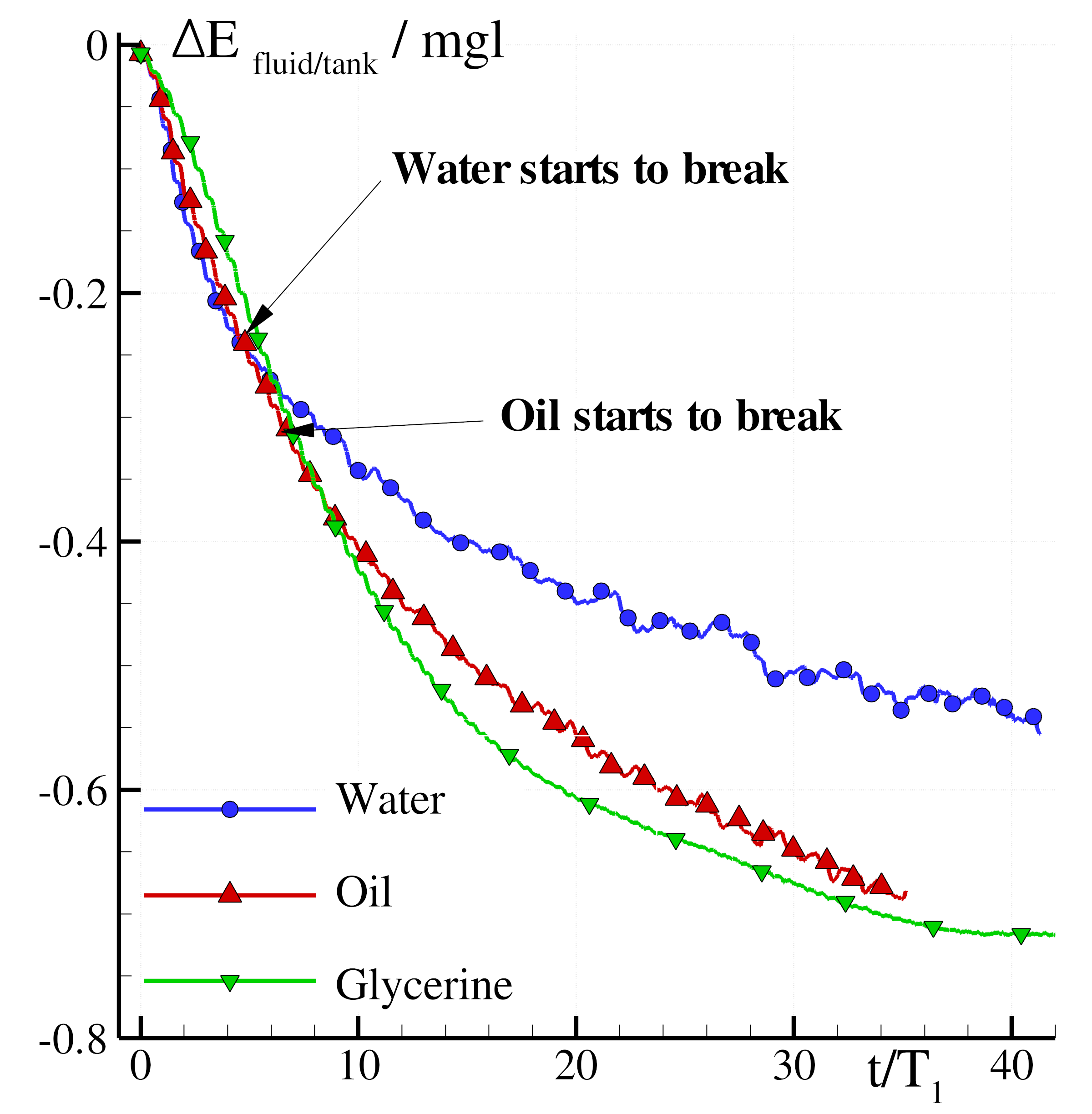}
\caption{Envelope function $\Phi$ (left) and energy transfer $\Delta E_{fluid/tank}$ (right) plotted as a function of time, with an excitation amplitude $A_m=0.10$ m,
using water, sunflower oil and glycerine inside the tank.
}
\label{fig:Delta_E_fluid_tank_oil_glycerine_am10}
\end{figure}

The left plot of Fig. \ref{fig:Delta_E_fluid_tank_oil_glycerine_am10} shows the
roll angle amplitude $\Phi$ plotted as a function of time for the amplitude $A_m = 0.10$ m using water, sunflower oil, and glycerine.
The attainment of the time-periodic state requires more oscillations than in
Series 1. Indeed, at least 50 periods are needed for all three liquids.
The experimental records are too short to take the system to the time-periodic state regime.
Experimental data can only be extrapolated upon in order to get the value of $\Phi$ for longer time ranges.
The time-periodic state angles are approximatively 12, 16 and 20 degrees for water, oil and glycerine respectively.
The least viscous fluid (water) leads to the lowest roll angles and the most viscous (glycerine)
to the largest ones as in the previous Series. However, the relative differences between the liquids are smaller than those observed in Series 1.

The right plot of Fig. \ref{fig:Delta_E_fluid_tank_oil_glycerine} depicts the
energy transfer $\Delta E_{fluid/tank}$ plotted as a function of time.
Considering the steepness of these curves during the initial stage,
it can be inferred that water has a faster response with respect to energy transfer.
Fig. \ref{fig:Delta_E_fluid_tank_oil_glycerine}
also reports the time instants when the first breaking of the free surface is observed.
Using water, the breaking appears earlier than with oil, that is, after just four periods.
When the breaking develops, the slope of $\Delta E_{fluid/tank}$
is reduced and the curve related to water crosses the curves corresponding to oil and glycerine.
In the early stages, the increase of $\Delta E_{fluid/tank}$ is affected by
the mechanical energy component (see eq. (\ref{eq:P_fluid_dissipation})). Water and oil
increase their mechanical energy quickly as a result of the tank motion. The breaking
process then induces a large increase of the dissipation component  $\Delta E_{fluid}^{dissipation}$;
it is the only one present at time-periodic state.
Due to the breaking processes for water and oil, and to the high viscosity of the glycerine,
the value of $\Delta E_{fluid/tank}$ increases one order of magnitude with respect to the measurements in Series 1.

\begin{figure}[ht!]
\centering
\includegraphics[width=0.90\textwidth]{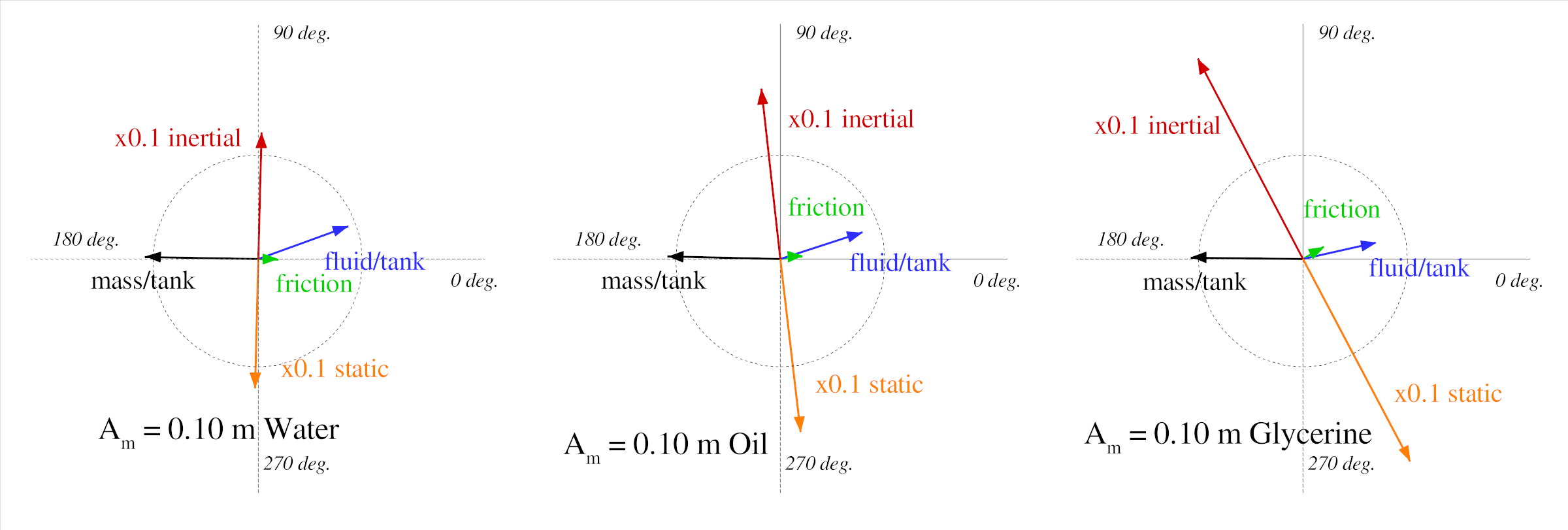}
\caption{Phasors obtained for the tank filled with water $A_m = 0.1 m$. \label{fig:Phasors_fluids}}
\end{figure}

The phasors of the various torques are plotted in Fig. \ref{fig:Phasors_fluids}. With water and oil, the phasor plane keeps close to the ideal TLD of Part I.
\begin{table}[ht!]
\begin{center}
    \begin{tabular}{|l||c|c|c|}
        \hline
        $Liquid$                     & water  & oil   & glycerine   \\ \hline \hline
        $\Phi$ [degree]              & 12     & 16    & 20          \\ \hline
        $\delta$ [degree]            & 85     & 96    & 120         \\ \hline
        $\Psi$ [degree]              & -60    & -80   & -110        \\ \hline
        $\Phi^{liquid}/\Phi^{empty}$ & 0.43  & 0.57  & 0.71       \\ \hline
        $\Delta E_{mass/tank}^{liquid}/\Delta E_{mass/tank}^{empty}$
                                     & $\sim 1.7$   & $\sim 2.3$  & $\sim 2.3$        \\ \hline
    \end{tabular}
\caption{\label{tab:oil_glycerine_am10} Tank filled with water, oil, glycerine: values of the main quantities reached at time-periodic state for the excitation amplitude $A_m$= 0.05 m.}
\end{center}
\end{table}

For this Series 2, an estimation of the main quantities at time-periodic state using
the three different liquids are reported in table \ref{tab:oil_glycerine_am10}.
The work done by the sliding mass on the tank, $\Delta E_{mass/tank}$, is very similar when using glycerine or oil, although the pair of values ($\delta$,$\Phi$) are different. The phase lags $\delta$ and $\Psi$ increase in modulus together with viscosity.


\subsection{Series 3} \label{sss:a015}

\begin{figure}[t!]
\centering
\includegraphics[width=0.3\textwidth]{Water_A150_A}
\includegraphics[width=0.3\textwidth]{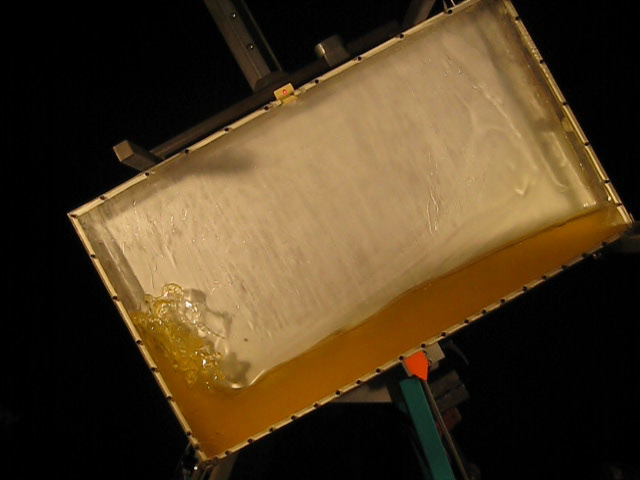}
\includegraphics[width=0.3\textwidth]{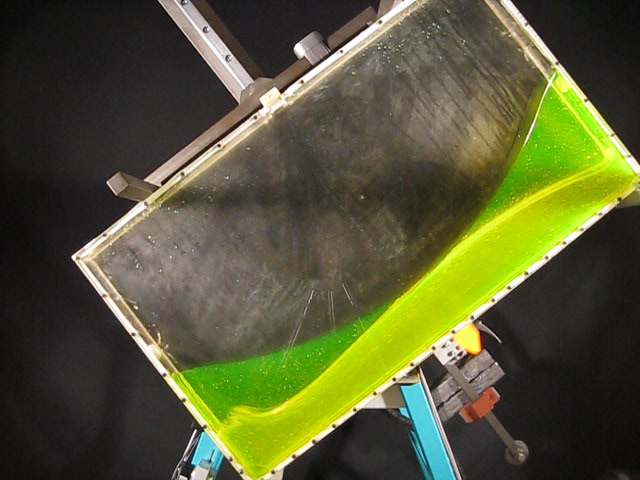}
\includegraphics[width=0.3\textwidth]{Water_A150_B}
\includegraphics[width=0.3\textwidth]{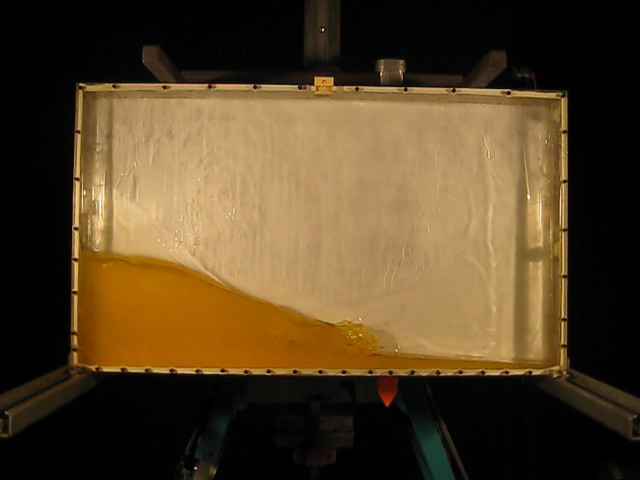}
\includegraphics[width=0.3\textwidth]{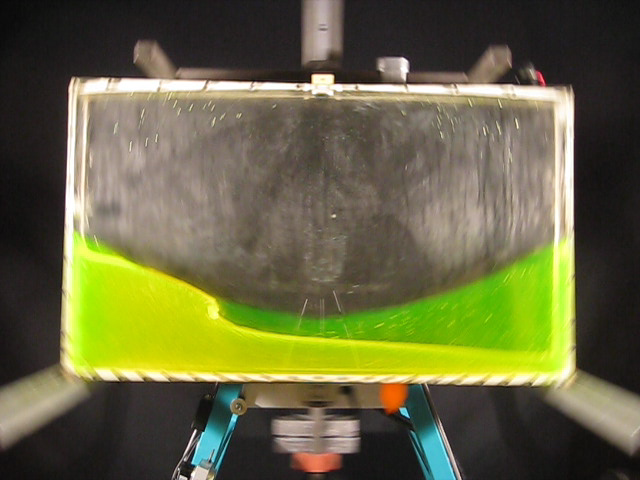}
\caption{Sloshing wave in tank:  filled with water, oil and glycerine $A_m = 0.15 m$. See supplementary materials from
[URL will be inserted by AIP].
} \label{fig:EvolSlosh_A150}
\end{figure}

The sloshing flows for the Series 3 cases are visualized
with several photographs (see Fig. \ref{fig:EvolSlosh_A150}).
The selected pictures correspond to an oscillation angle of about $30^\circ$,
which is extremely large for roll motion sloshing.
With water, almost the entire side wall is covered, and the successive evolution shows a large plunging wave with a splash-up
producing air entrapment. For the oil, the elevation on the wall is similar, whilst the breaking wave is less violent.
With the glycerine, a spilling breaker develops, a salient feature to be found in a liquid 740 times more viscous than water.

\begin{figure}[ht]
\centering
\includegraphics[width=0.400\textwidth]{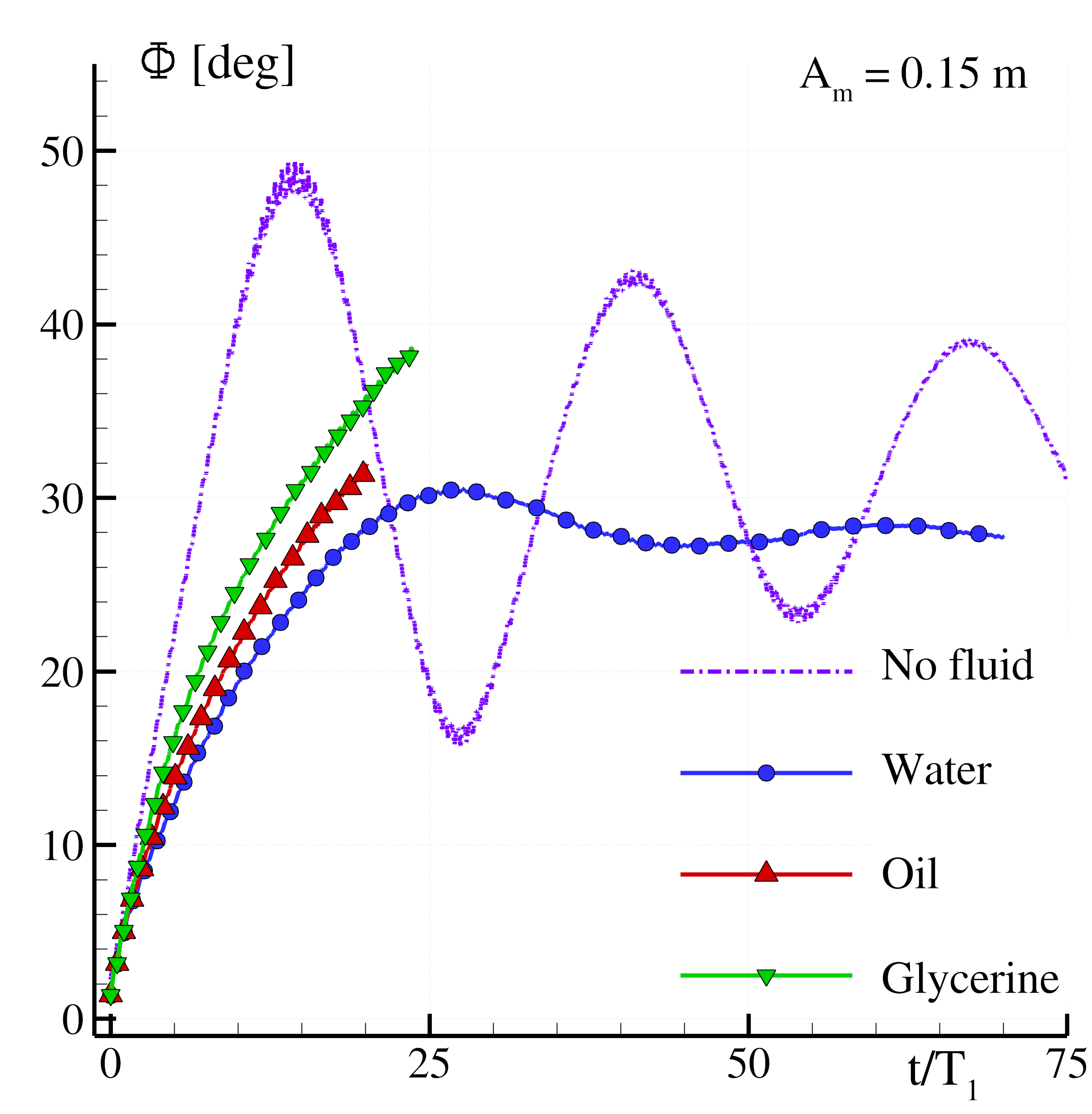}
\includegraphics[width=0.400\textwidth]{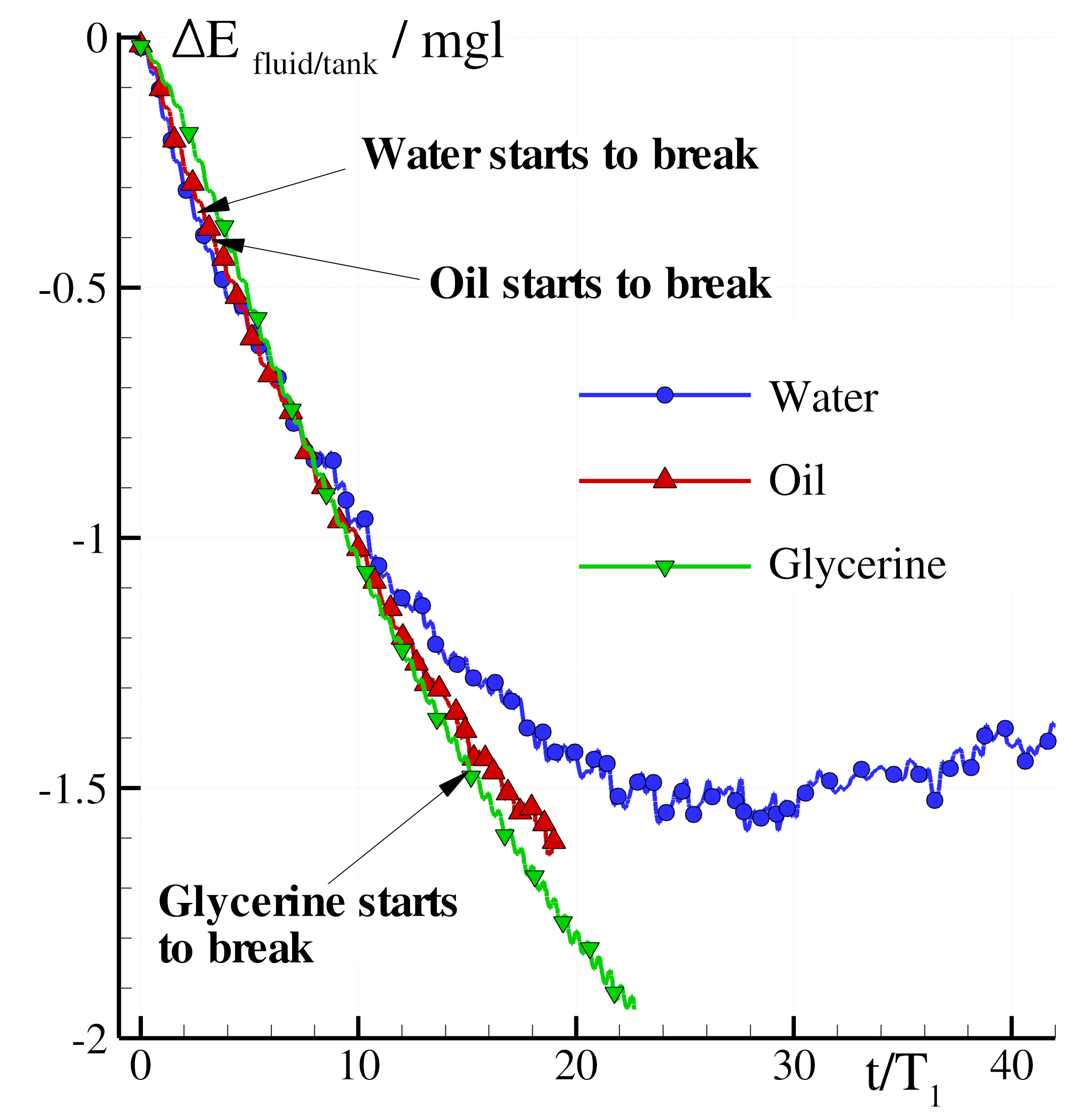}
\caption{Envelope function $\Phi$ (left) and energy transfer $\Delta E_{fluid/tank}$ (right) plotted as a function of time, with an excitation amplitude $A_m=0.15$ m,
using water, sunflower oil and glycerine inside the tank.
}
\label{fig:Delta_E_fluid_tank_oil_glycerine_am15}
\end{figure}

The left plot of Fig. \ref {fig:Delta_E_fluid_tank_oil_glycerine_am15} shows the time histories for the envelope
function $\Phi$. When oil and glycerine are used, the roll angle exceeds $35^\circ$
and the experiment is stopped in order to avoid risking the integrity of the rig.
This reduces the possibilities of the analysis, but the time history of $\Phi$ suggests that with oil and glycerine the roll angle at time-periodic state can be higher than the one reached with the empty tank.

The large roll motion induces an intense energy transfer $\Delta E_{fluid/tank}$ (see right plot of Fig.
\ref{fig:Delta_E_fluid_tank_oil_glycerine_am15}), almost three times larger than in Series 2.
The time instants when the breaking phenomena start to develop are reported in Fig. \ref{fig:Delta_E_fluid_tank_oil_glycerine_am15}. The first breaking event occurs during the first four periods of oscillations for oil and water. Glycerine shows an extended
transient phase; the first breaking event appears after 15 periods.

\begin{figure}[ht]
\centering
\includegraphics[width=0.9\textwidth]{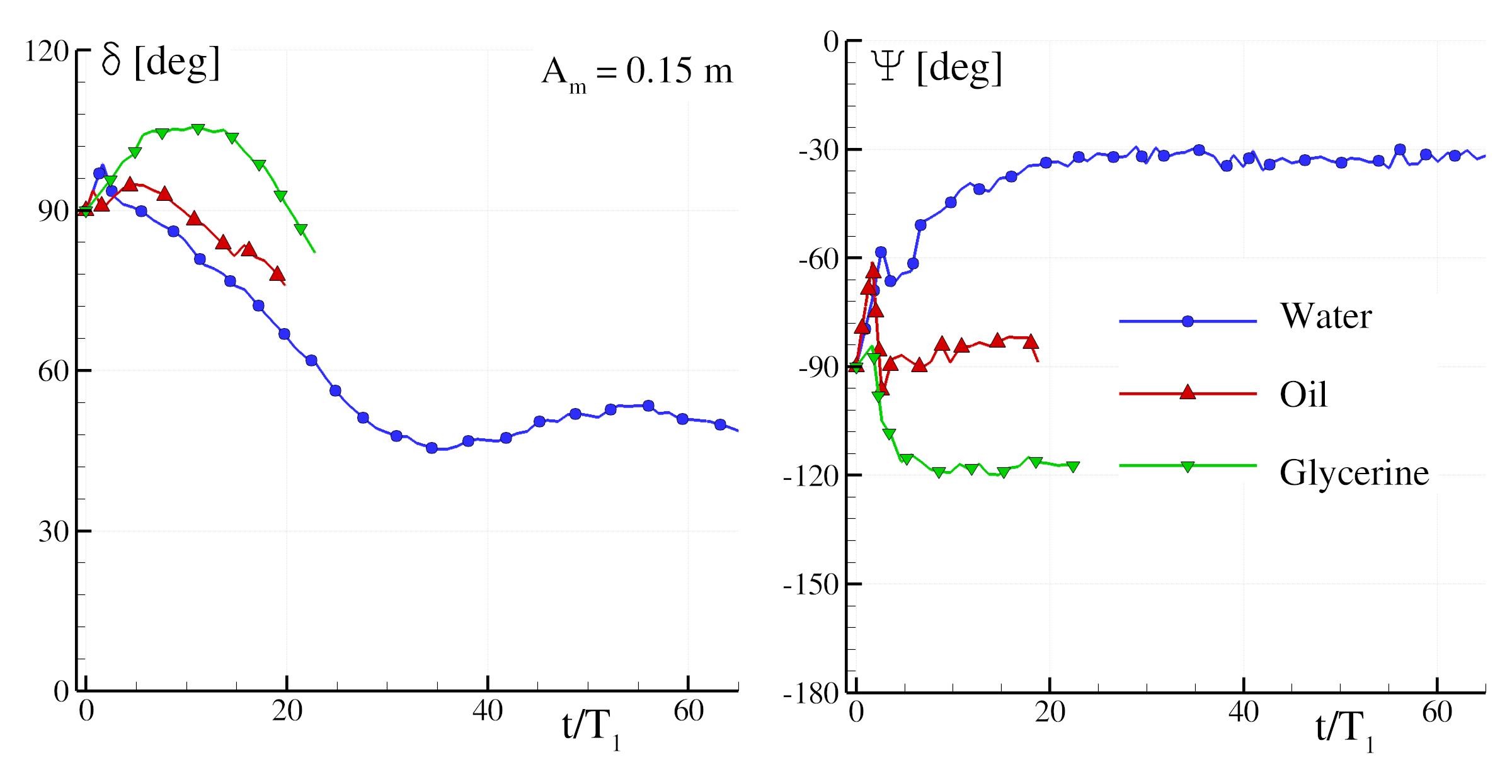}
\caption{Phase lags $\delta$ (left) and $\Psi$ (right) plotted as a function of time, with an excitation amplitude $A_m=0.15$ m (Series 3),
using water, sunflower oil and glycerine.
}
\label{fig:Delta_Psi_oil_Glycerine_Am015}
\end{figure}

Fig. \ref{fig:Delta_Psi_oil_Glycerine_Am015} shows the phase lags $\delta$ and $\Psi$ plotted as a function of time for Series 3. Large time ranges are needed to stabilize the phase lag $\delta$ (left panel) between the sliding mass and the roll angle.
However, even if the dynamical system is far from a time-periodic state, $\Psi$ (right panel) is almost stable after 20 periods.
The increase of viscosity generates an increase in modulus for both $\delta$ and $\Psi$,
at least in the time range covered by the experiments, similar to what was obtained in previous Series.
These results also show that when using oil in the tank, the torque $M_{fluid/tank}$ is in quadrature with the roll motion. This affects the energy transfer between the fluid and the tank. Indeed, the
$\Delta E_{fluid/tank}$ obtained with oil is almost the same as the one obtained with glycerine, even though the roll angle with oil is lower.


\subsection{Series 4} \label{sss:a020}

Very little information is available for the last Series, mainly because after a few periods the roll angles induced
are larger than allowed by the rig, and therefore the experimental time records are too short for a complete analysis
of the system's dynamics.

\begin{figure}[ht!]
\centering
\includegraphics[width=0.40\textwidth]{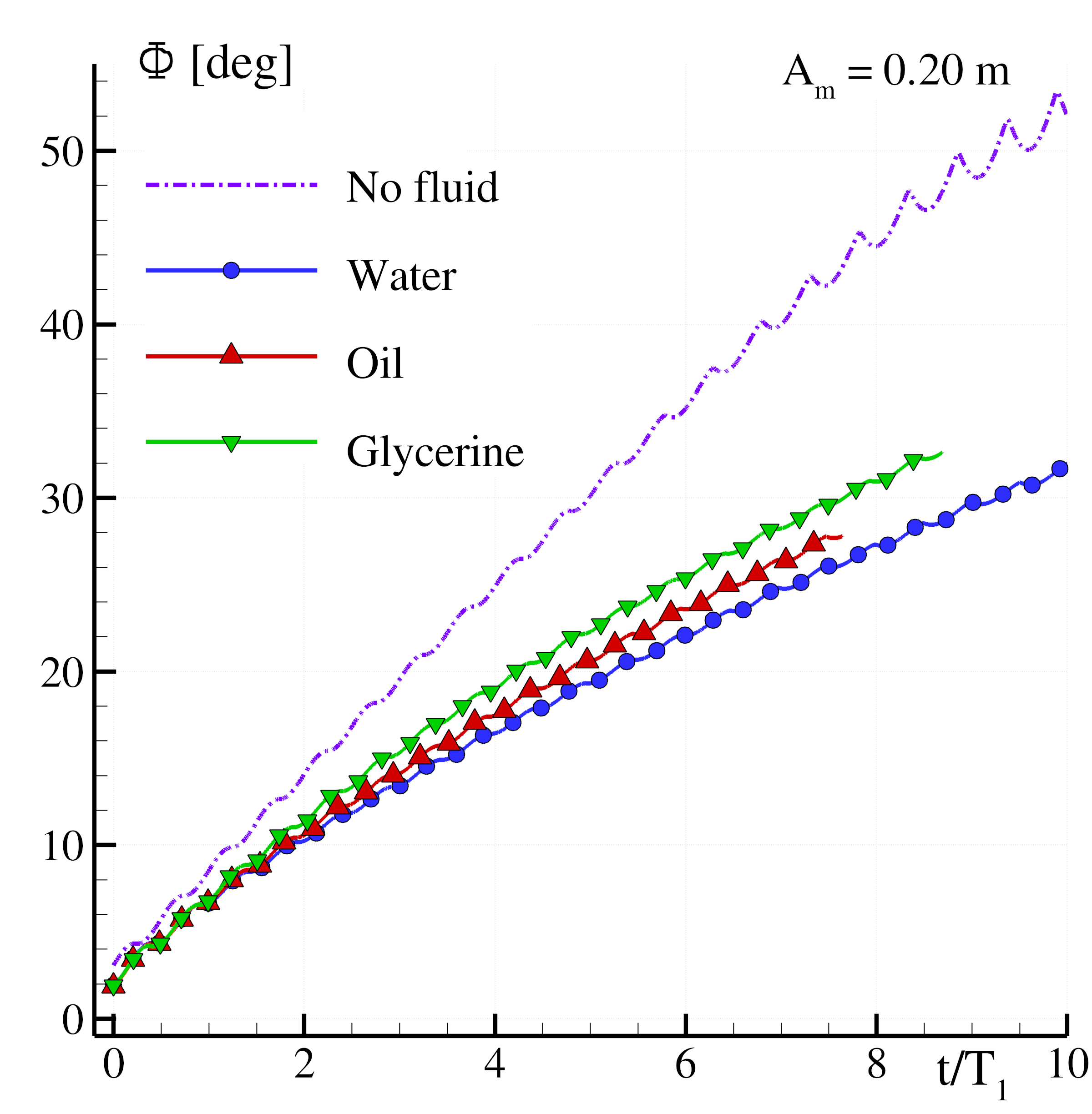}
\includegraphics[width=0.400\textwidth]{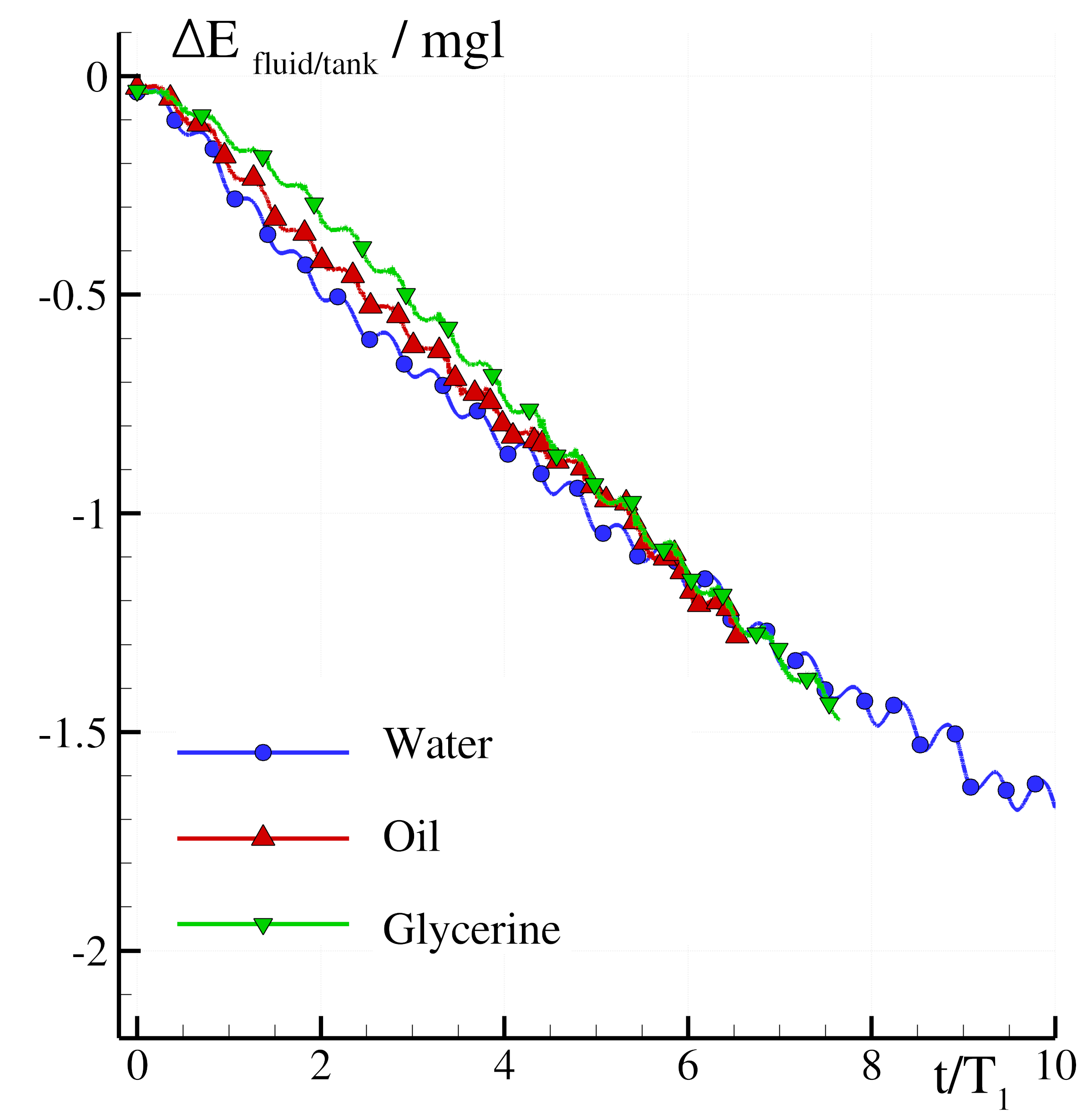}
\caption{Envelope function $\Phi$ (left) and
energy transfer $\Delta E_{fluid/tank}$ (right) plotted as a function of time, with an excitation amplitude $A_m=0.20$ m,
using water, sunflower oil and glycerine.
}
\label{fig:Delta_E_fluid_tank_oil_glycerine_am20}
\end{figure}

\begin{figure}[ht!]
\centering
\includegraphics[width=0.9\textwidth]{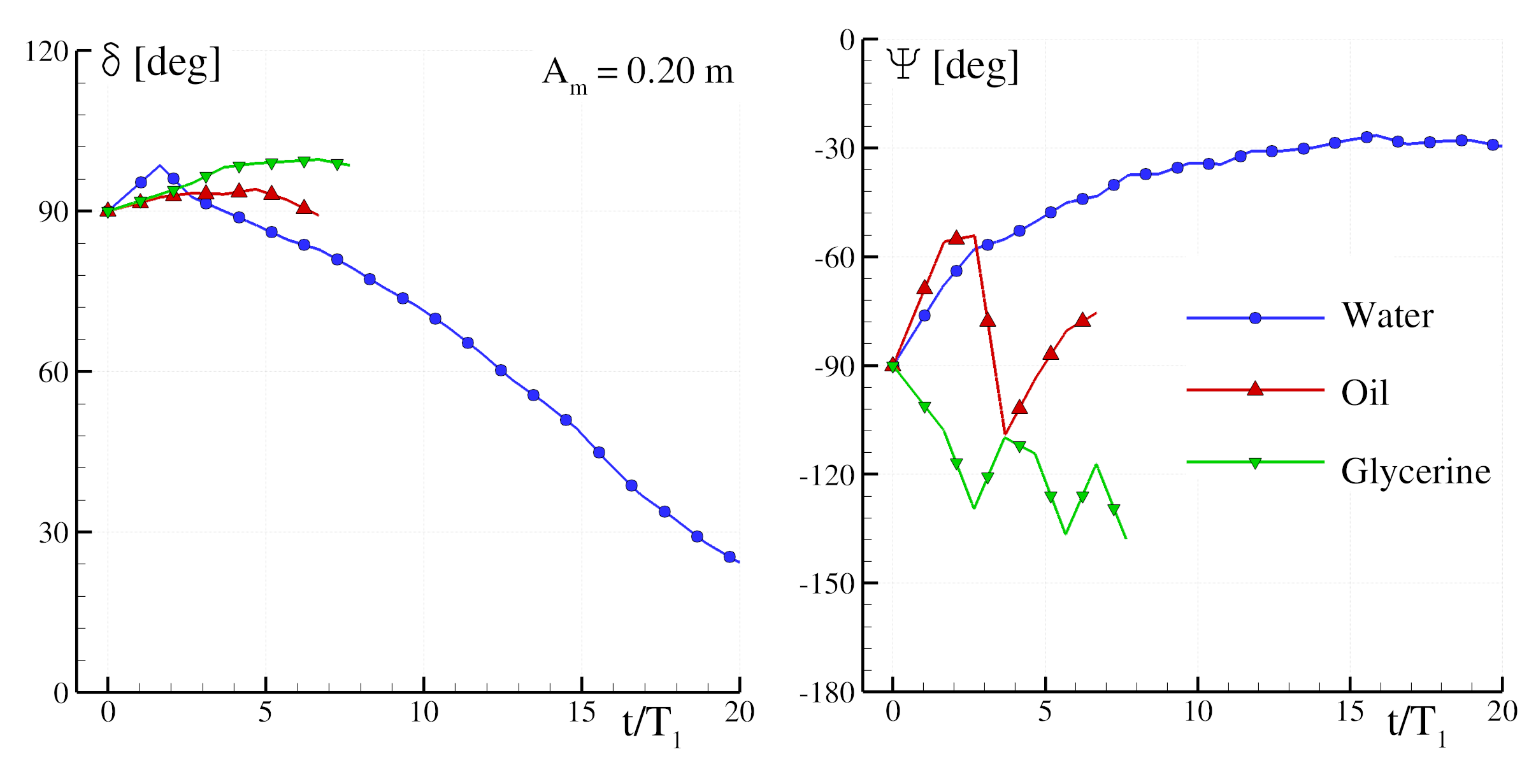}
\caption{Phase lags $\delta$ (left) and of
$\Psi$ (right) plotted as a function of time, with an excitation amplitude $A_m=0.20$ m,
using water, sunflower oil and glycerine.
}
\label{fig:Delta_Psi_oil_Glycerine_Am020}
\end{figure}

The left plot of Fig. \ref{fig:Delta_E_fluid_tank_oil_glycerine_am20} shows the envelope function for the roll angle $\Phi$
obtained for the first eight periods. The angles relative to the three different liquids manifest small differences. Again,
as already shown for the first three Series, the roll angle attained is smallest when using water.
The energy transfer $\Delta E_{fluid/tank}$ is also very similar for the three liquids in the transient stage.
In these first eight oscillations, the phase lag $\delta$ remains close to $90^\circ$ (left plot of Fig. \ref{fig:Delta_Psi_oil_Glycerine_Am020}) whilst the phase lag $\Psi$ has a more complex time behavior.
When using oil, $\Psi$ shows a tendency to remain closer to a quadrature configuration $90^\circ$ whilst, when using water, $\Psi$ decreases in modulus until the value of $-30^\circ$.
Contrarily, when using glycerine, $\Psi$ increases in modulus, finally settling on the value $-120^\circ$.

\begin{figure}[ht!]
\centering
\includegraphics[width=0.9\textwidth]{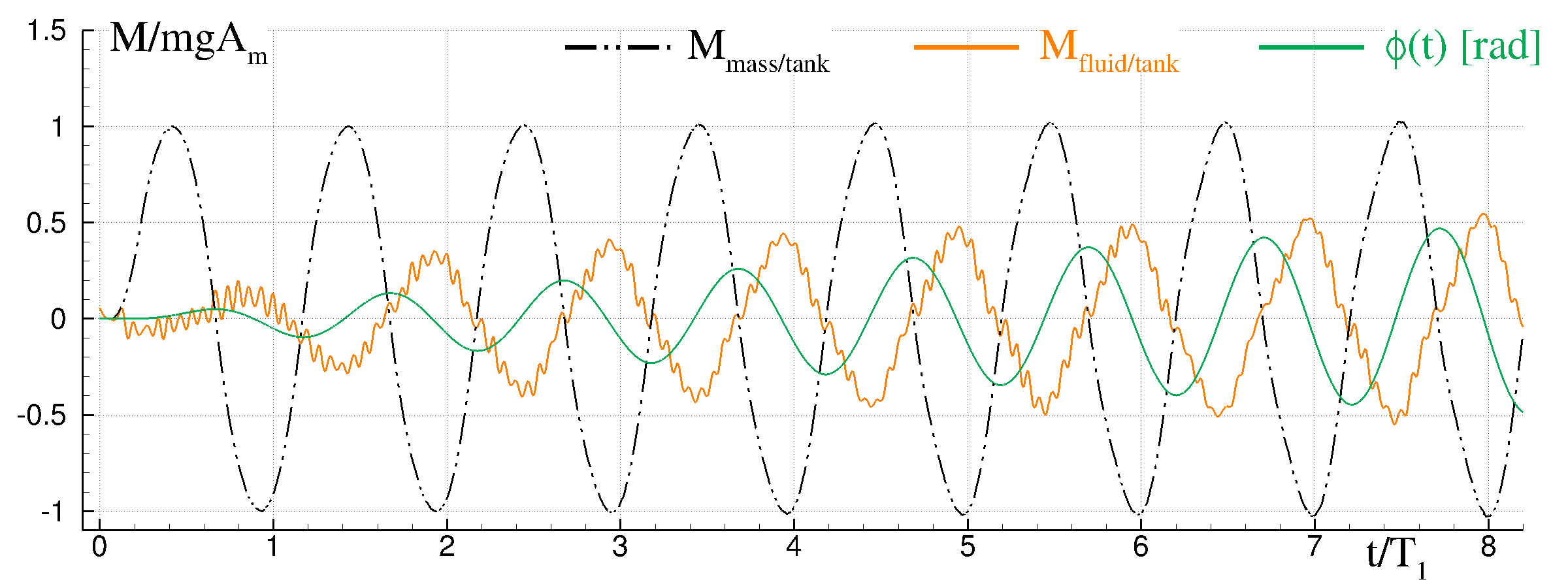}
\caption{Torque $M_{mass/tank}$ and $M_{fluid/tank}$ as a function of time
recorded during the first period of oscillation with an excitation amplitude $A_m=0.20$ m and
using sunflower oil.
}
\label{fig:M_fluidTank_Oil_Am020}
\vskip 0.2cm
\end{figure}

Fig. \ref{fig:M_fluidTank_Oil_Am020} depicts
$M_{mass/tank}$, $M_{fluid/tank}$ and $\phi$ in the transient stage,  as a function of time.
$M_{fluid/tank}$ quickly opposes $M_{mass/tank}$ ({\emph i.e.} $\delta+\Psi\approx 0$).
Conversely $M_{fluid/tank}$ increases in amplitude very slowly, indicating that, in the first 10 periods
of oscillation, the system is quite far from the time-periodic state, as confirmed by the increase of the roll
angle $\phi(t)$. Furthermore, $M_{mass/tank}$ shows a small non-linear behavior resulting from the large amplitude $A_m$,
as already commented upon in section \ref{ss:water}.
%
%
\subsection{Summary of dissipation results for the different liquids}\label{Summarydissipation}
\begin{figure}[hb!]
\centering
\includegraphics[width=0.32\textwidth]{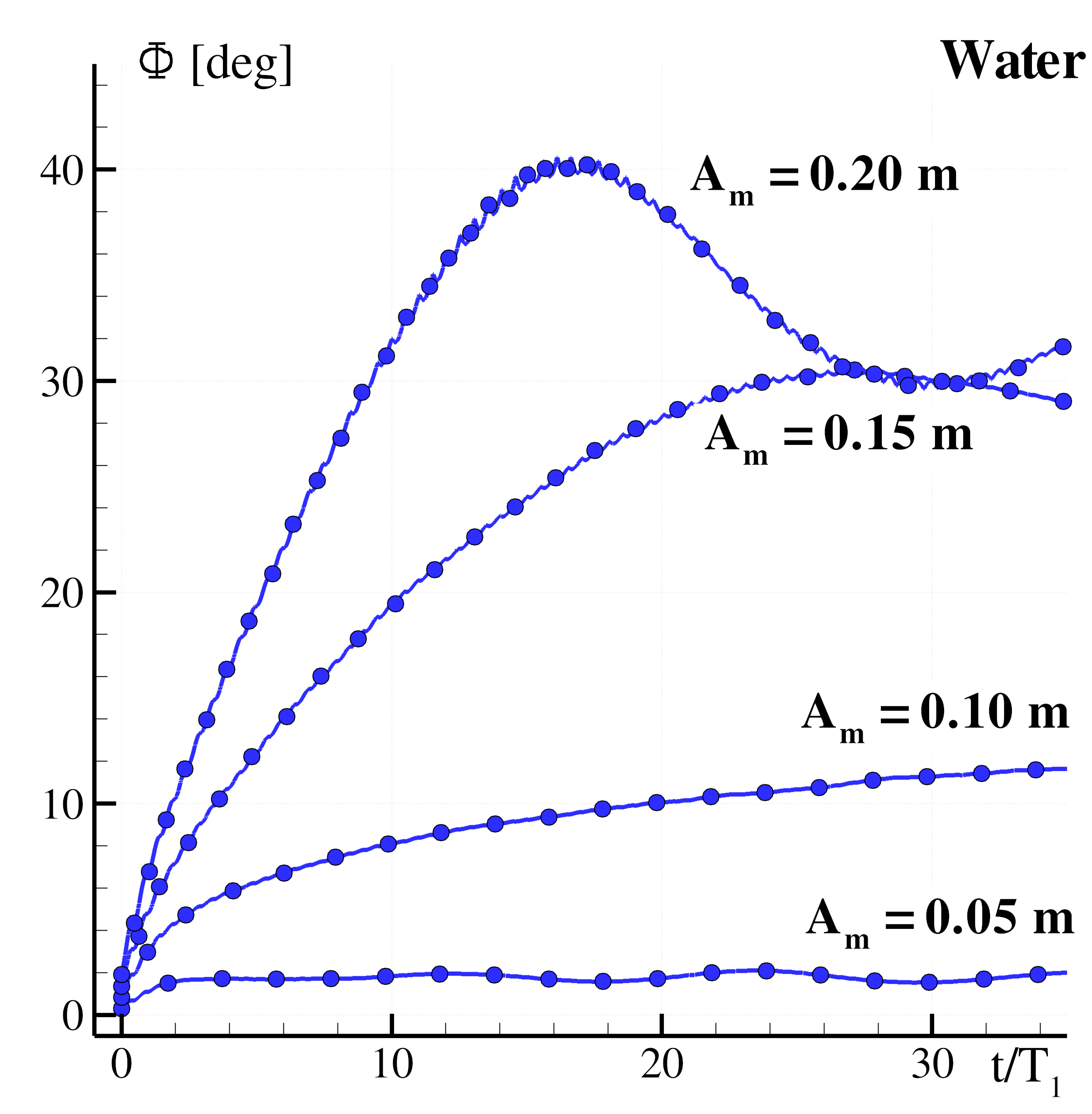}
\includegraphics[width=0.32\textwidth]{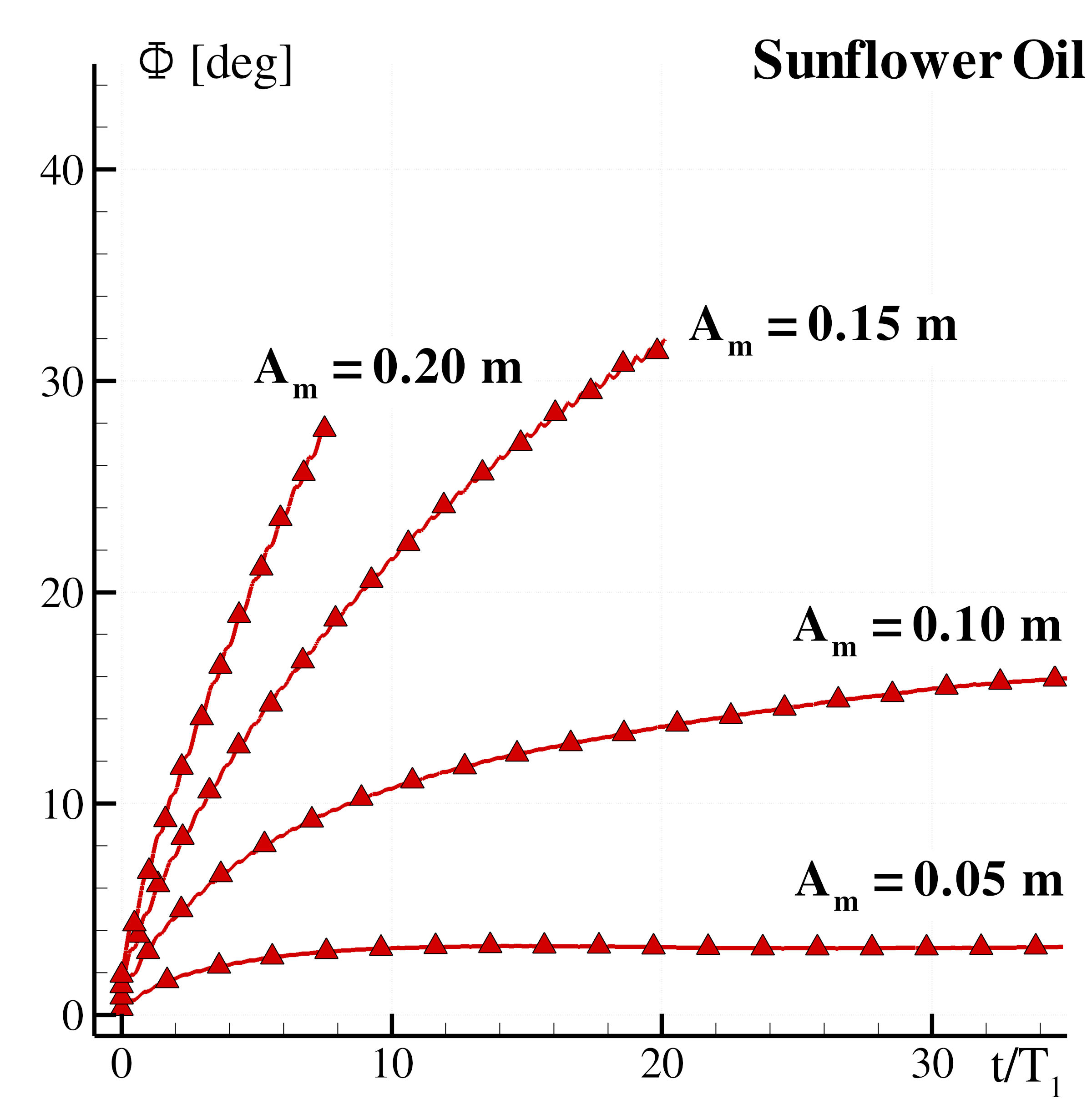}
\includegraphics[width=0.32\textwidth]{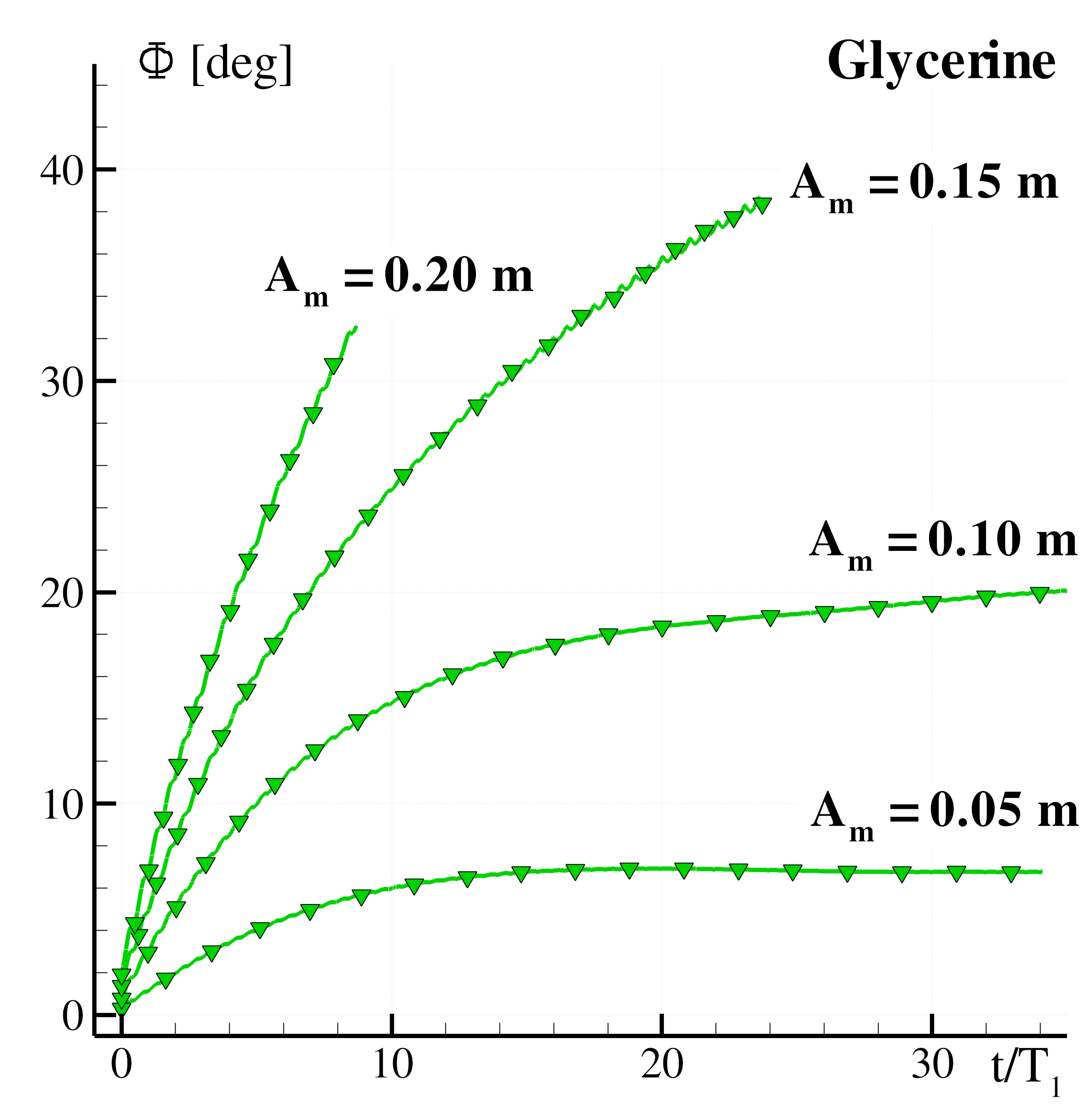}
\vskip 0.2cm
\includegraphics[width=0.32\textwidth]{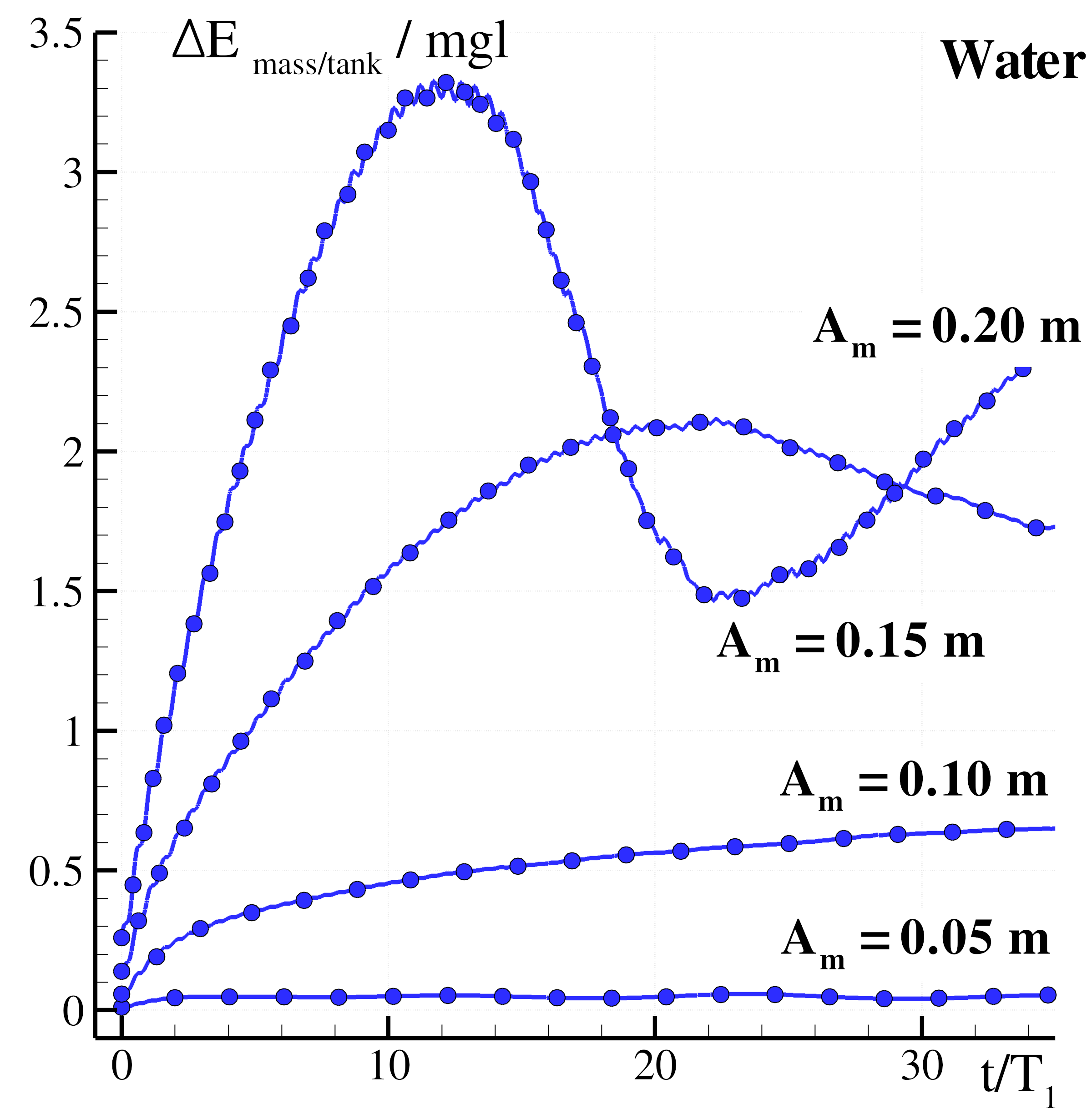}
\includegraphics[width=0.32\textwidth]{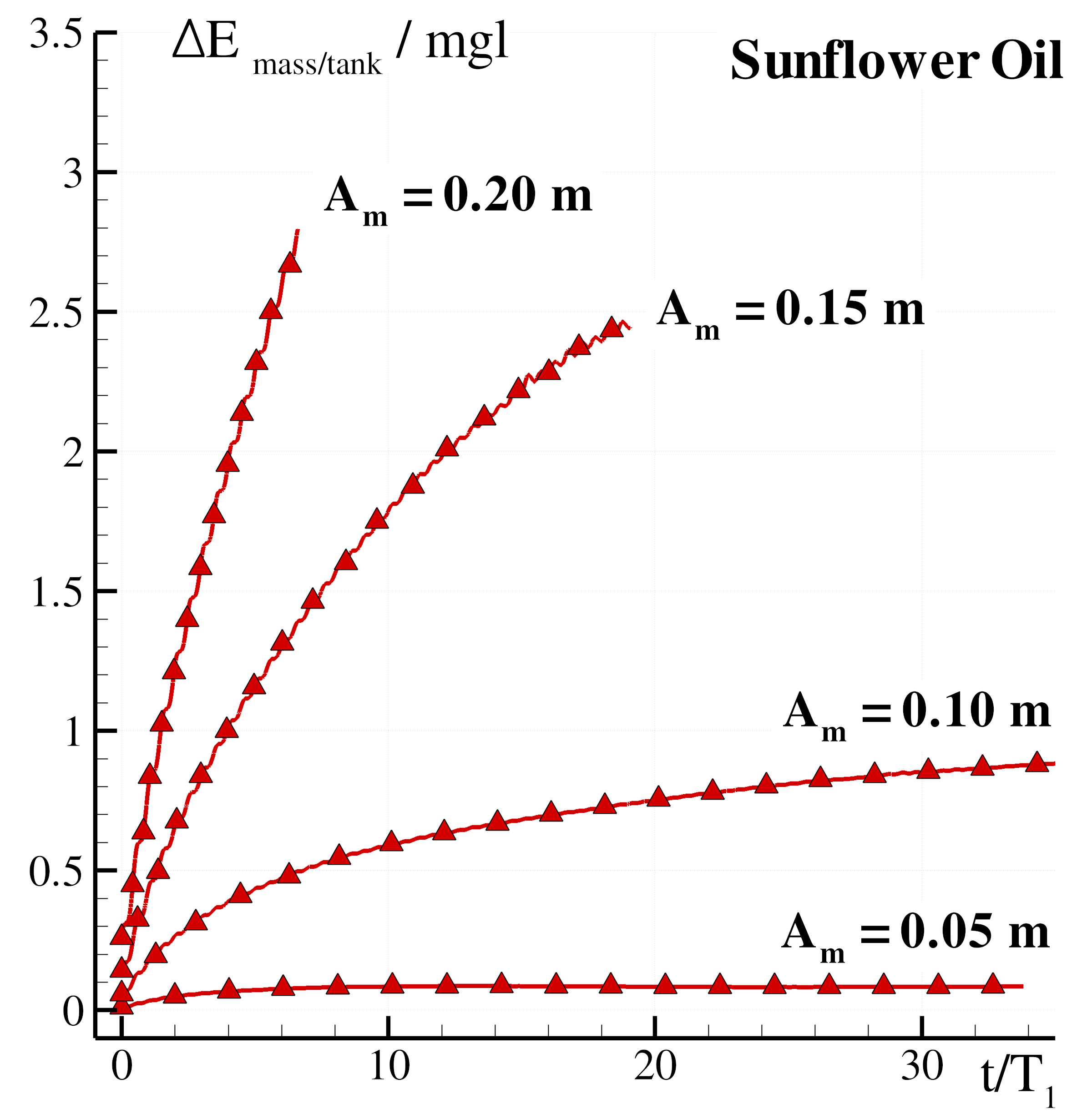}
\includegraphics[width=0.32\textwidth]{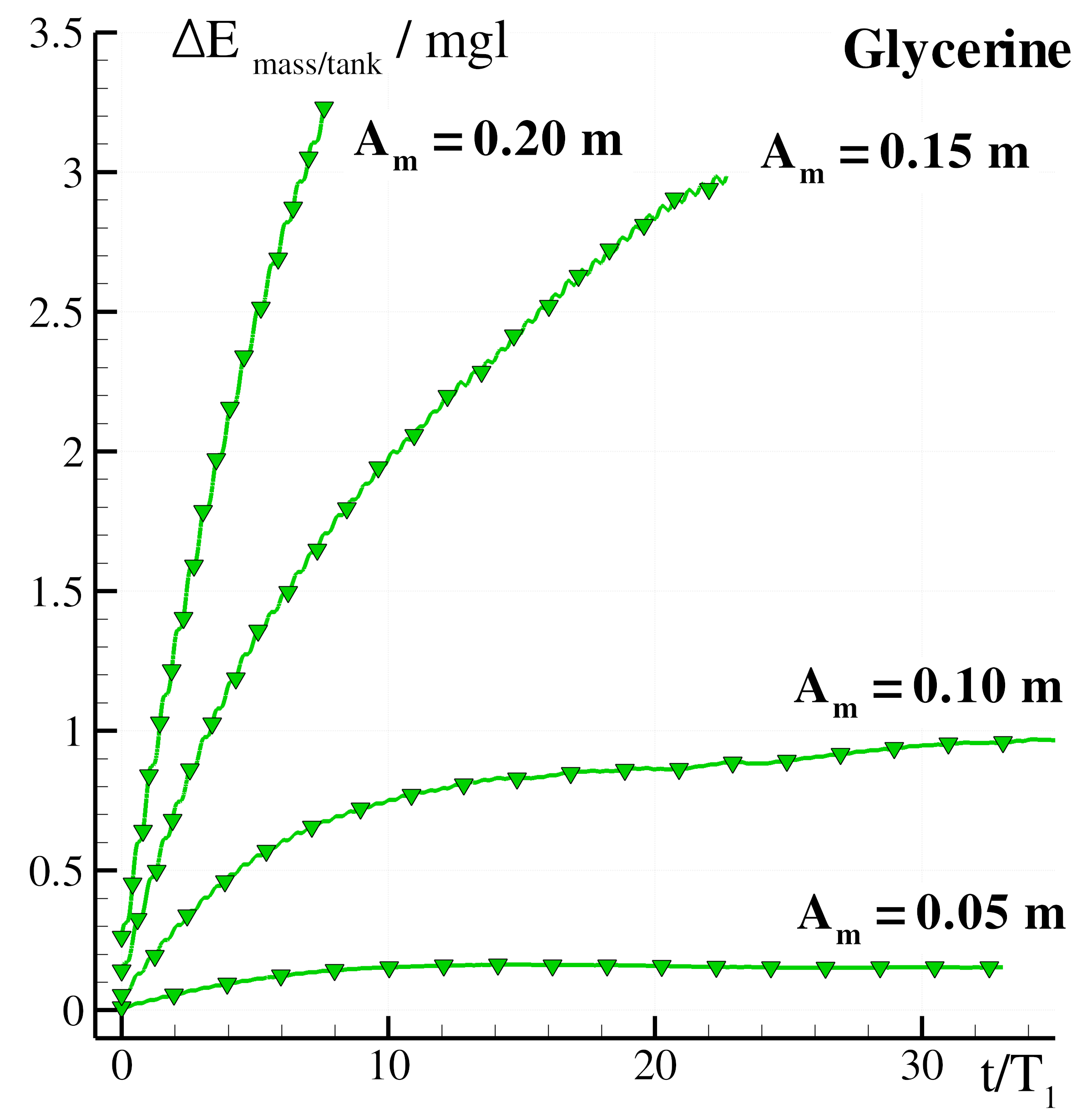}
\caption{Roll angle $\phi(t)$ (top) and of energy transfer $\Delta E_{mass/tank}$ (bottom) plotted as a function of time
for all the excitation amplitudes, varying the liquid present inside the tank.
}
\vskip 0.1cm
\label{fig:Phi_DE_mass_tank_all_Am}
\end{figure}

Putting together all the data recorded in the four Series, some conclusions
can be drawn about the influence of dissipation
induced by sloshing on the system's dynamics.
Fig. \ref{fig:Phi_DE_mass_tank_all_Am} shows the time histories for $\Phi$ and $\Delta E_{mass/tank}$
for the four Series using the three different liquids.
According to this figure, increasing the viscosity of the liquid drives both $\Phi$ and $\Delta E_{mass/tank}$ to increase.

However, a more detailed analysis can be achieved for the performance of the sloshing liquid
using the theoretical findings introduced in sections IV  
and V 
of Part I and the evaluation of the viscosity coefficient $\alpha$ (see equation (\ref{Energy_SPH_Verhagen}) expected to be close to the theoretical value 1.68.

\begin{figure}[t!]
\centering
\includegraphics[width=0.32\textwidth]{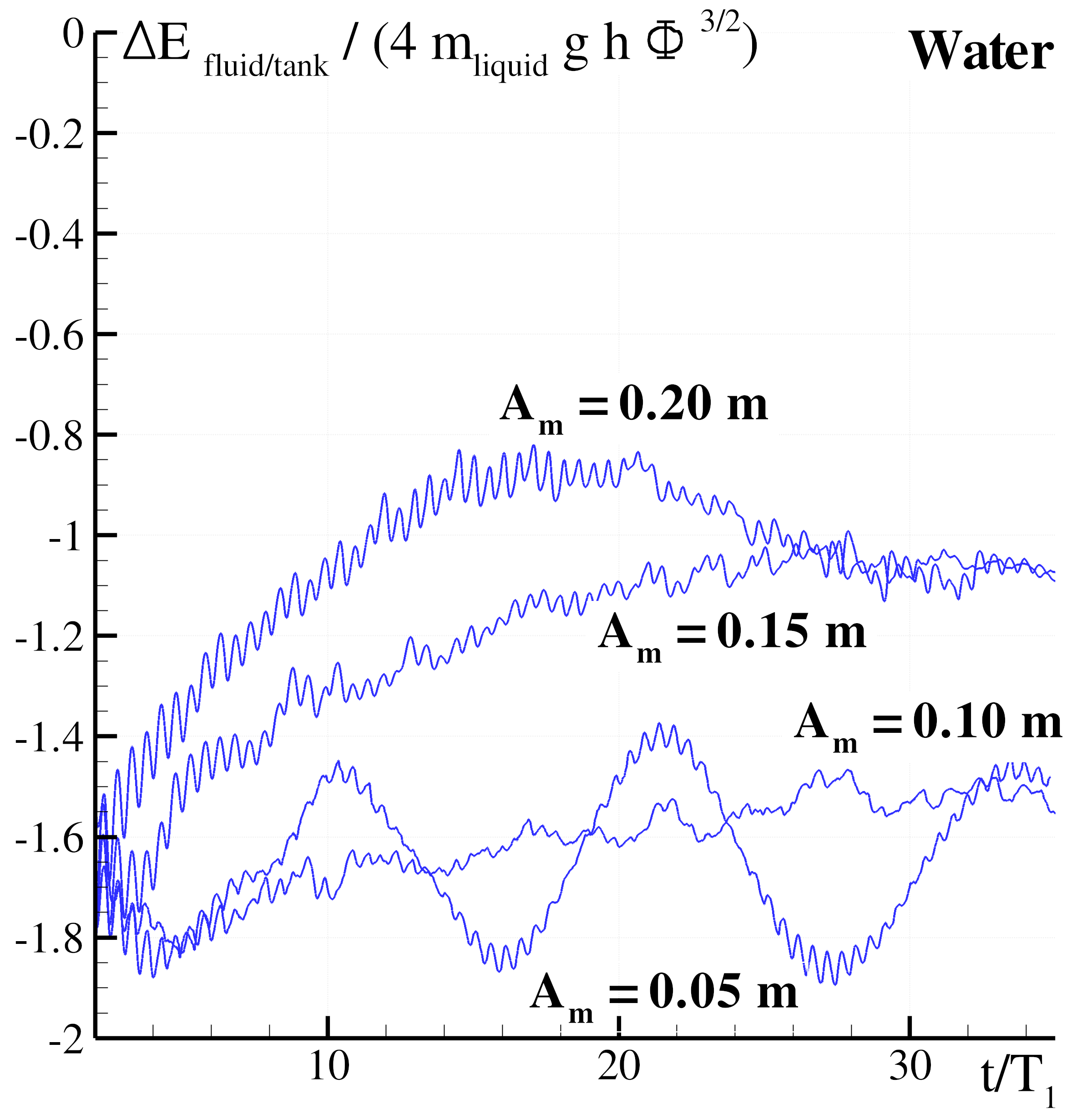}
\includegraphics[width=0.32\textwidth]{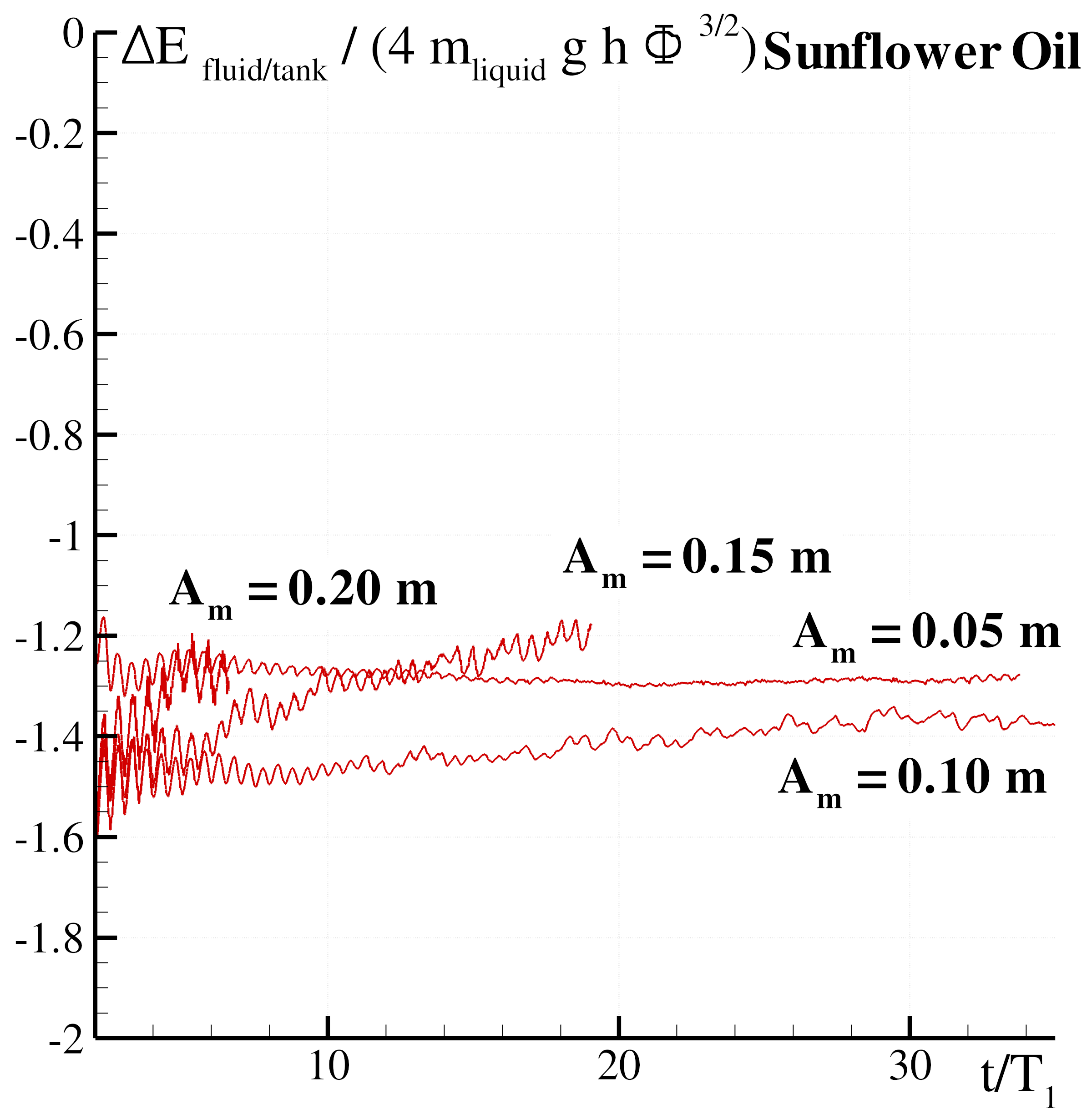}
\includegraphics[width=0.32\textwidth]{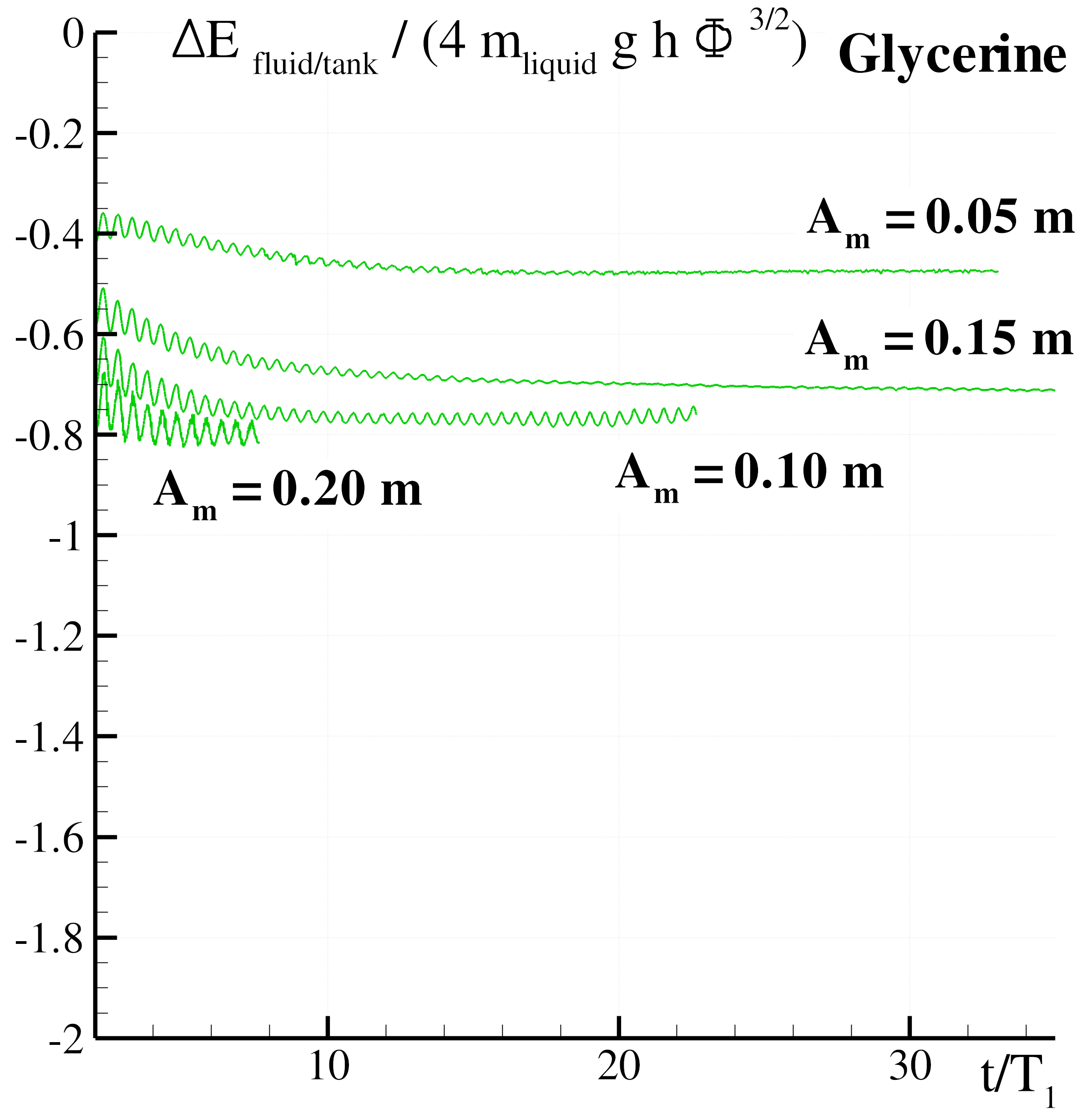}
\caption{Energy transfer $\Delta E_{fluid/tank}$ plotted as a function of time for all the excitation amplitudes, using water (left), sunflower oil (middle) and glycerine (right) inside the tank.
}
\label{fig:DE_fluid_tank_all_Am}
\end{figure}

Fig. \ref{fig:DE_fluid_tank_all_Am} shows the energy transfer $\Delta E_{fluid/tank}$ for the four Series obtained experimentally. This energy is made non dimensional using the coefficient $(4\,m_{liquid}\,g\,h\,\Phi^{\frac{3}{2}})$. Since the energy transfer between the fluid and the tank,  $\Delta E_{fluid/tank}$, is fully dissipative at a time-periodic state, the ratio $\Delta E_{fluid/tank}/(4\,m_{liquid}\,g\,h\,\Phi^{\frac{3}{2}})$ is expected to converge to $-\alpha$ over time.

The left plot of Fig. \ref{fig:DE_fluid_tank_all_Am} concerns the Series with water. For the first two
Series, the $\alpha$ coefficient oscillates around the theoretical value
but tends to diminish when $A_m$ increases. This means that for the most energetic case, the
sloshing flow decreases its dissipation efficiency.
Sunflower oil shows a similar behavior, but with much smaller variations.
Glycerine shows a more regular behavior but  this time  the coefficient $\alpha$ is
very low with respect to the theoretical one. This can be linked to the fact that its high viscosity inhibits
the formation of hydraulic jumps and their associated dissipation.
Furthermore, $\alpha$ rises with the glycerine only when increasing the excitation amplitude $A_m$.
Even if the glycerine presents a smaller $\alpha$ value, this is partially compensated by its larger density, which, in absolute terms, induces a certain increase on $\Delta E_{fluid}^{dissipation}$.

\begin{figure}[t!]
\centering
\includegraphics[width=0.99\textwidth]{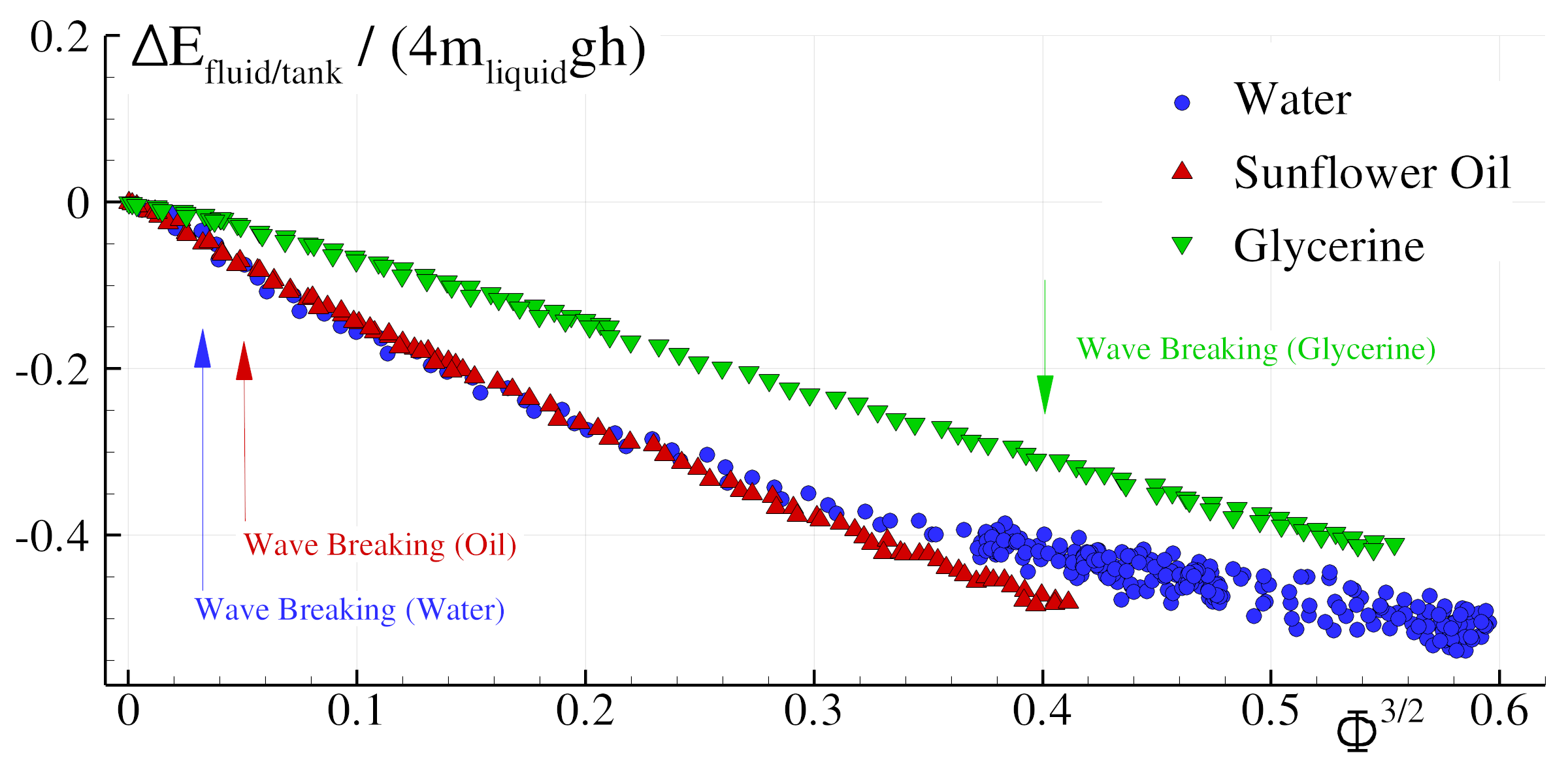}
\caption{Energy transfer $\Delta E_{fluid/tank}$ plotted as a function of $\Phi^{3/2}$, using water, sunflower oil and glycerine.
}
\label{fig:Energy_Phi32}
\end{figure}

Another way to read the above results is to directly plot the experimental results
on the $[\Delta E_{fluid/tank}$ vs $\Phi^{3/2}]$ graph.
Fig. \ref{fig:Energy_Phi32} shows this same result with the energy made non-dimensional by the factor $4 m_{liquid}\,g\,h$.
On this plot the time dependence disappears and
points related to the transitory stages are plotted together with
those related to the periodic conditions.
This confirms that $\Delta E_{fluid/tank}$ is essentially a dissipative
contribution for the sloshing conditions studied in this work.
The data related to all three liquids lie on
curves where the steepness is linked to the $\alpha$ coefficient.
As shown previously for water, $\alpha$ decreases as the roll angle $\Phi$
increases. Water and sunflower oil exhibit a similar behavior.
Considering the mass $m_{liquid}$ of the three liquids (see table
\ref{fluid_matrix}) $\Delta E_{fluid/tank}$ is comparable for large $\Phi$.

\section{Practical application}
From the analysis performed in previous sections, it can be seen that the effective
roll angle $\Phi$ reached by the system  depends not only on the amplitude of the
sliding mass motion but on the phase lags $\delta$ and $\Psi$,
which in turn depend on the fluid, the filling height $h$, and the excitation frequency $\omega$.

In Part II of this paper series, the analysis has been limited to one filling height and one frequency
of excitation $\omega=\omega_1$.
The analysis performed also shows that the energy dissipation capability with this system can be large.
Indeed, the system first transforms the external energy given by the sliding mass into
mechanical energy, making the tank move. Then, the large roll motion induces
violent sloshing and its associated dissipation
which prevent the energy accumulated by the tank to be sent back to the sliding mass.

\begin{figure}[ht!]
\centering
\includegraphics[width=0.5\textwidth]{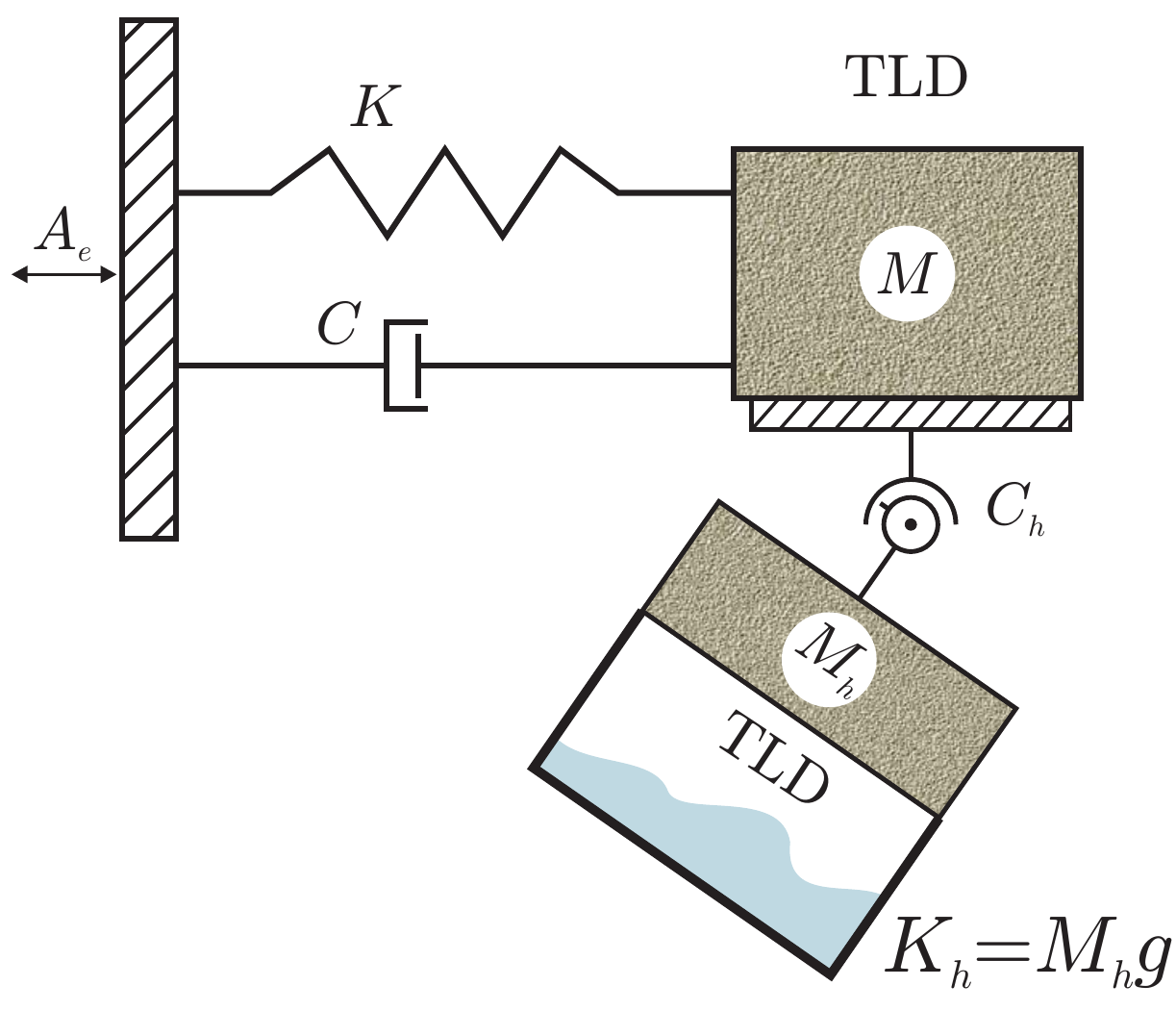}
\caption{Mechanical sketch of the use of the pendulum has a hybrid pendulum mass liquid damper (HPMLD).}
\label{fig:HPMLD}
\end{figure}

Considering this analysis and noting that the forcing term in a
horizontally driven pendulum \citep{lourenco2011design}
is analogous to the forcing term of the present system (equation (II.4) of Part I),  
a novel damping system is proposed. It is named here a hybrid pendulum mass liquid damper (HPMLD).
The mechanical sketch of such a system is shown in Fig. \ref{fig:HPMLD}.
Its principle is that of a HMLD, (see section II.B in Part I), 
but the secondary damping system is a pendulum partially filled with liquid.

The idea to transfer energy from horizontal motions into pendular
ones (with a mass and no liquid) was already presented by \citet{tedesco1999structural},
and a full analysis of a such a device, a Pendulum Tuned Mass Damper (PTMD),
is carried out by \citet{lourenco2011design}. For an analysis of an oscillating structure
containing liquid and coupled with a horizontal motion device, see the work of \citet{Pirner2007}.

The system proposed herein incorporates both ideas. Its capability to dissipate large
amounts of energy through sloshing and wave breaking has been discussed in this paper.
An in-depth analysis of its full potential as a HPMLD is still left for future work.

%
\section{Conclusions}

A Pendulum-TLD system has been analyzed in terms of its kinematics, dynamics and energy dissipation mechanisms.

It is composed of three coupled sub-systems: a sliding mass, the moving part of an angular motion sloshing rig, including an empty tank, and the fluid which partially fills that tank. The data from a set of experiments in which the sliding mass is excited with harmonic motions is discussed.

The experimental analysis of this system has been performed at the mechanical system resonance frequency. This choice together with the filling height adopted, both stem from Part I of the present paper series. The experimental analysis of other frequencies and other filling heights is left for future work. The test matrix has included fluids of different densities and viscosities, namely water, sunflower oil and glycerine, and four different exciting amplitudes.

The kinematics and dynamics of the flow are complex and extremely nonlinear. Low amplitude traveling waves
occur for the small excitation cases, while large breaking waves are found for large excitations.

The different physical quantities can be evaluated with great accuracy since
the experimental set-up allows for the computation of all terms from the roll angle time history and its derivatives.
In particular, the time evolution of energy balances between the fluid and the tank can be evaluated.
At time-periodic state this balance becomes the energy dissipated by the fluid.

The experiments have been analyzed considering modulus and phase for the different torques acting on the system. The phase lags between the sliding mass motion, roll angle, and torque exerted by the fluid allows for the understanding of the time-periodic state reached by the system in terms of roll amplitude and work done by the sliding mass.

The fluid dissipation measured during the experiments has been compared to the theoretical model described in Part I. The predictions obtained with this model match those corresponding to the water cases.
Since the energy dissipation in such a model is obtained though a hydraulic jump solution, the conclusion is that the dissipation source for these cases is mostly due to breaking.

The influence of the nature of the liquid on the system has been studied for a set of excitation amplitudes. At the time-periodic state, the most viscous liquid, glycerine, leads to the largest oscillation amplitude. On the contrary, using the least viscous liquid, water, leads to the smallest oscillation amplitude. The liquid choice is shown to greatly influence the phase lags between the different torques.
During the early transient stage, the least viscous fluid (water) is shown to have a faster increase of the roll counteracting effect, and to reach a time periodic state within a smaller number of periods.

When correctly scaled using the theoretical model, the fluid dissipation value for a fixed roll angle diminishes as viscosity increases.
However, since the time-periodic state roll angle is larger when increasing viscosity, the net dissipation is also larger. This is another conclusion from this research and it implies that the choice of liquid when obtaining the best damping performance is not straightforward.

For the empty tank condition, the system may return energy to the sliding mass; this does not happen when fluid
is present, even when similar angles are found at time-periodic state. This idea inspires a
novel damping device, named hybrid pendulum mass liquid damper (HPMLD), which is introduced here.
This system may be able to take advantage of the intense mechanical response of the structure in order to pump the initial energy and later dissipate it with the violent sloshing flow generated. Its proper development and analysis is left for future work.



%
\section*{Acknowledgements}
The research leading to these results has received funding from
the Spanish Ministry for Science and Innovation under
grant TRA2010-16988 ``\textit{Ca\-rac\-te\-ri\-za\-ci\'on Num\'erica y
Experimental de las Cargas Fluido-Din\'amicas en el transporte de Gas Licuado}'' .

This work has been also funded by the Flagship Project RITMARE - The Italian Research for
the Sea - coordinated by the Italian National Research Council and funded by the Italian Ministry of
Education, University and Research within the National Research Program 2011-2013.

The authors are grateful to Sonny Mendez and Hugo Gee for English language proofreading.
\bibliographystyle{apsrev}
\bibliography{bib}

\end{document}